\documentclass[a4paper,fleqn,usenatbib]{mnras}

\usepackage{newtxtext,newtxmath}
\usepackage[T1]{fontenc}
\usepackage{ae,aecompl}

\usepackage{graphicx}	% Including Fig. files
\usepackage{amsmath}	% Advanced maths commands
\usepackage{mathtools}
\usepackage{amssymb}	% Extra maths symbols
\usepackage{natbib}
\usepackage{hyperref}
\usepackage{multirow}
\usepackage{dcolumn}
\usepackage{array}
\usepackage[colorinlistoftodos]{todonotes}

\newcolumntype{L}[1]{>{\raggedright\let\newline\\\arraybackslash\hspace{0pt}}m{#1}}
\newcolumntype{C}[1]{>{\centering\let\newline\\\arraybackslash\hspace{0pt}}m{#1}}
\newcolumntype{R}[1]{>{\raggedleft\let\newline\\\arraybackslash\hspace{0pt}}m{#1}}

\mathchardef\mhyphen="2D
\newcommand{\tsc}[1]{_{\textsc{#1}}}
\newcommand{\mrm}[1]{{{\mathrm{#1}}}}
\newcommand{\bs}[1]{\boldsymbol{#1}}

\newcommand{\pow}[1]{\times 10^{#1}}
\newcommand{\sn}[5]{\textsc{S-}{#1}\_\textsc{V-}{#2}\_\textsc{K-}{#3}\_\textsc{R-}{#4}\_\textsc{O-}{#5}}
\newcommand{\sns}[2]{(#1-#2)}
\newcommand{\rc}{r\tsc{core}}
\newcommand{\tff}{t\tsc{ff}}
\newcommand{\kmin}{k\tsc{min}}
\newcommand{\avir}{\alpha\tsc{vir}}
\newcommand{\fbes}{f\tsc{bes}}
\newcommand{\Ms}{\, \mrm{M}_{\odot}}
\newcommand{\pc}{\, \mrm{pc}}
\newcommand{\gcm}{\, \mrm{g} \, \mrm{cm}^{-3}}
\newcommand{\kms}{\mrm{km} \, \mrm{s}^{-1}}
\newcommand{\comment}[1]{}

\title[Episodic outflows in protostellar cores]{The impact of episodic outflow feedback on stellar multiplicity and the star formation efficiency}

\author[Rohde et al.]{
P. F. Rohde,$^{1}$\thanks{E-mail: rohde@ph1.uni-koeln.de}
S. Walch,$^{1}$
S. D. Clarke, $^{1}$
D. Seifried,$^{1}$
A. P. Whitworth,$^{2}$\\
\newauthor{ and A. Klepitko$^{1}$} \\
\\
$^1$ \, I. \, Physikalisches \, Institut, \, Universit\"at \, zu \, K\"oln, \, Z\"ulpicher \, Str. \, 77, \, D-50937 \, K\"oln, \, Germany\\
$^2$ \, School \, of \, Physics \, and \, Astronomy, \, Cardiff \, University, \, Cardiff \, CF24 \, 3AA, \, UK\\
}

\date{Accepted XXX. Received YYY; in original form ZZZ}

\pubyear{2020}

\begin{document}
\label{firstpage}
\pagerange{\pageref{firstpage}--\pageref{lastpage}}
\maketitle

%%%%% [Words=219]
\begin{abstract}
The accretion of material onto young protostars is accompanied by the launching of outflows. Observations show that accretion, and therefore also outflows, are episodic. However, the effects of episodic outflow feedback on the core-scale are not well understood. We have performed 88 Smoothed Particle Hydrodynamic simulations of turbulent dense $1 \Ms$ cores, to study the influence of episodic outflow feedback on the stellar multiplicity and the star formation efficiency (SFE). Protostars are represented by sink particles, which use a sub-grid model to capture stellar evolution, inner-disc evolution, episodic accretion and the launching of outflows. By comparing simulations with and without episodic outflow feedback, we show that simulations with outflow feedback reproduce the binary statistics of young stellar populations, including the relative proportions of singles, binaries, triples, etc. and the high incidence of twin binaries with $q\geq 0.95$; simulations without outflow feedback do not. Entrainment factors (the ratio between total outflowing mass and initially ejected mass) are typically $\sim 7\pm 2$, but can be much higher if the total mass of stars formed in a core is low and/or outflow episodes are infrequent. By decreasing both the mean mass of the stars formed and the number of stars formed, outflow feedback reduces the SFE by about a factor of 2 (as compared with simulations that do not include outflow feedback). 
\end{abstract}

%%%%%
%%%%%

\section{Introduction}
One of the fundamental open questions in modern astrophysics is why molecular gas is very inefficiently converted into stars. On molecular cloud scales, the star formation efficiency (SFE) is only a few percent \citep{Leroy08, Utomo18, Schruba19}, and stellar feedback is presumed to be the reason for this low efficiency \citep{Murray11}. Along with stellar winds, ionising radiation and supernovae, protostellar outflows are one of the feedback mechanisms that might substantially reduce the overall SFE, particularly in regions where there are no massive stars \citep[e.g.][]{Nakamura07, Hansen12, Federrath14, Krumholz14, Li18, Cunningham18}. 

Low-mass stars form preferentially in prestellar cores, which tend to be concentrated in dense filaments inside molecular clouds \citep{Shu87, Andre07, Myers09, Andre14, Konyves15, Marsh16, Konyves20}. In contrast to the elongated shapes of filaments, prestellar cores are approximately spherical, and their density profiles are often described as Bonnor--Ebert spheres, with typical radii of $R\tsc{core} \sim 0.01\,\pc$ to $0.1\,\pc$ \citep{Bonnor56, Ebert57, Johnstone00, Alves01, Tafalla04, Konyves20}. The prestellar core mass function (CMF) approximates to a log-normal distribution with a peak around $\sim 0.5 \Ms$ \citep{Konyves15,Konyves20,Marsh16}. Molecular line observations of prestellar cores show non-thermal velocity components indicating internal turbulence \citep{Andre07, Pineda11, Rosolowsky17}.

Protostellar outflows often accompany the star formation process \citep{Bally16}. Observations and numerical simulations suggest that outflows consist of two components: a collimated high-velocity jet \citep{Mundt83, Reipurth01, Tafalla10, Lee17} and a slower wide-angle disc-wind, launched further out in the accretion-disc \citep{Machida14, Tabone17, Liu18, Louvet18, Zhang19}. Both components are known to be rotating \citep{Lee17, Hirota17, Zhang18, Zhang19}. Therefore, outflows carry away angular momentum from the disc-star system, which in turn allows the central protostar to accrete while staying below its break-up speed \citep{Pudritz07, Bjerkeli16}.

Protostellar jets are launched from the innermost regions of protostellar accretion discs. Numerous authors have simulated protostellar outflows self-consistently using magneto-hydrodynamic simulations \citep[e.g.][]{Hennebelle11, Machida09, Seifried12, Price12, Machida13, Machida14, Bate14, Tomida14, Tomida15, Lewis17, Machida19, Saiki20}.  However, such simulations must resolve the launching region down to $r_{\textsc{launch}} \sim  \mathrm{R}_{\odot}$ to reproduce the extremely high-velocity jet component that originates in the inner-most disc region. It is presently not computationally feasible to follow the evolution of protostars through the whole protostellar phase using such a high resolution. Other authors therefore mitigate this problem by invoking almost resolution-independent sub-grid models to launch outflows \citep{Nakamura07, Cunningham11, Peters14, Myers14, Federrath14, Offner14, Kuiper15, Offner17, Li18, Rohde18}.

Exactly how the gas is launched is still not well understood  \citep[see, e.g. the reviews of][]{Arce07, Frank14, Bally16}. However, the consensus is that outflows are accretion powered: gravitational energy is converted into kinetic and magnetic energy, which then drives and collimates the outflow, either through magnetic pressure or magneto-centrifugal forces \citep{Blandford82, Konigl00, Lynden-Bell03, Pudritz07, Machida08, Seifried12}. Since the accretion onto a protostar is episodic, outflows are also episodic \citep{Reipurth89, Hartigan95, Hartmann97, Konigl00, Arce07, Hennebelle11, Kuiper15, Bally16, Choi17, Cesaroni18, Samal18, Zhang19}. 

Protostellar outflows inject a significant amount of energy and momentum into the surroundings \citep{Arce10, Plunkett13, Feddersen20}, and are likely to have a profound impact on their host cores. This is especially true in the context of low-mass star formation where other feedback mechanisms do not come into play. The `primary' ejected gas from the immediate vicinity of the protostar entrains `secondary' core material, thereby carving out a cavity which widens over time \citep{Arce06}. Within the cavity, accretion flows onto the protostar are suppressed, lowering the amount of gas which can fall directly onto the protostar, and hence lowering the protostellar accretion rate. \citep{Wang10}. The resulting feedback loop of accretion and outflow launching is not fully understood. Because outflows act to disperse a star's birth core, they are presumed to play a role in terminating the accretion process \citep{Zhang16}. Theoretical studies show that this may cause the SFE on core scales to be as low as 15-50\% \citep{Machida13, Offner14, Offner17}. However, more observations and theoretical studies are needed to fully understand the effects of outflow feedback on core scales.

The stellar initial mass function (IMF) \citep{Kroupa2002, Chabrier03} and the initial statistics of multiple systems \citep[e.g.][]{Duchene13} are key constraints on theories of star formation. \cite{Raghavan10} find that in the field roughly 50\% of systems are single stars like our Sun; all the rest are binaries or higher-order multiples (i.e. triples, quadruples, quintuples, etc., hereafter HOMs). Recent observations have started to reveal the multiplicity statistics of pre-main-sequence stars \citep{Duchene07, Connelley08, Chen13, Pineda15, Tobin16, Shan17, Duchene18, Tobin18, Kounkel19}. \cite{Tobin16} have observed the Perseus molecular cloud using the VLA and report an overall multiplicity fraction of $m\!f = 0.4$ for Class 0/I protostars. Like  \cite{Chen13}, they find that $m\!f$ decreases for later evolutionary stages. Dynamical N-body interactions are probably the main reason for the decay of HOMs \citep[e.g.][]{Bate2005a, Goodwin07}. Another observed property of low-mass stellar multiples is the excess of almost equal-mass binary systems, referred to as `twin' binaries \citep{Lucy06, Simon09, Kounkel19}. 

Although multiplicity statistics are an important benchmark for simulations of star formation, such simulations should also reveal the detailed physical processes that deliver the observed multiplicity statistics \cite[see, e.g.][]{Offner11, Bate12, Lomax15, Li18, Kuffmeier19, Wurster19b}. Here we explore the effect of outflow feedback on the formation and evolution of multiple systems.

The paper is structured as follows. In Section~\ref{SEC:Method}, we describe the computational method, outline modifications to the sub-grid outflow model developed earlier by \citet{Rohde18}, and define the initial and boundary conditions. In Section~\ref{SEC:Results}, we present the results of the simulations and discuss how the stellar properties depend on the initial conditions. In Section~\ref{SEC:Multiplicity}, we describe the multiplicity statistics and how they are influenced by outflow feedback. In Section~\ref{SEC:SelfReg}, we analyse the properties of the outflows and their relation to the SFE. In Section~\ref{SEC:Conclusion} we summarise our results.

%%%%%
\section{Computational Method}\label{SEC:Method}
%%%%%

%%%%%
\subsection{SPH code Gandalf}\label{SEC:Gandalf}
%%%%%

For the hydrodynamical simulations, we use the highly object-orientated smoothed particle hydrodynamics (SPH) and mesh-less finite volume (MFV)  code \textsc{Gandalf} \citep{Hubber18}. {\sc Gandalf} adopts the `grad-h' SPH formulation \citep{Springel02} with an $\textsc{M}4$ kernel \citep{Monaghan85} and $\eta = 1.2$, giving on average $\sim 58$ neighbours. {\sc Gandalf} uses hierarchical block time-stepping. In our simulations the number of allowed timestep levels is $N\tsc{lvl} = 9$. Therefore an SPH particle on the highest level has $2^{N\tsc{lvl}} = 512$ times more timesteps than a particle on the lowest level. During a timestep, all particles are allowed to adapt to higher levels if this is necessary. {\sc Gandalf} uses the artificial viscosity prescription proposed by \cite{Morris97}, regulated by a time-dependent switch \citep{Cullen10}. {\sc Gandalf} offers various integration schemes, and we choose the second-order Leapfrog KDK scheme.

As in \cite{Rohde18}, we use the approximate radiative heating and cooling algorithm of \cite{Stamatellos07}. This method uses local SPH particle quantities to estimate a mean optical depth, which is then used to compute heating and cooling rates. The method accounts for changes in specific heat due to dissociation and ionisation of H and He. The opacity accounts for ice-mantle evaporation and dust sublimation, as well as the switch from dust opacity to molecular-line opacity. In contrast to \cite{Stamatellos07}, we do not use the local gravitational potential, but the local pressure gradient, to estimate the mean optical depth. This change to the original method has been proposed by \cite{Lombardi15}, and improves the behaviour in non-spherical geometries, such as accretion discs and collision interfaces.  

%%%%%
\subsection{Sink particles}\label{SEC:SinkParticles}
%%%%%

Sink particles, as originally proposed by \cite{Bate95}, are used in prestellar core collapse simulations to limit the otherwise continuously decreasing timesteps. We use the improved sink particle description introduced by \cite{Hubber13}. Sink particles have radius $R\tsc{sink}\sim 1\,{\rm AU}$, and are introduced at densities exceeding $\rho\tsc{sink} = 10^{-10} \, \gcm$. We use gravitational softening on scales of order $R\tsc{sink}$ to make the N-body integration more robust. SPH particles in the vicinity of a sink particle are not accreted instantaneously, but smoothly over a few timesteps. Therefore the vicinity of a sink particle is not empty, and this leads to improved hydrodynamical behaviour. Besides limiting the timesteps, sink particles serve as active star particles, each hosting the four sub-grid models detailed in the next four subsections.

%%%%
\subsection{Episodic accretion}\label{SEC:EpisodicAccretion}
%%%%

Following \cite{Stamatellos12a} we divide sink particles into an unresolved inner accretion disc (IAD) and a central protostar. We keep track of the masses,  $M\tsc{iad}$ and $M_{\star}$, and the angular momenta,  $\bs{L}\tsc{iad}$ and $\bs{L}_{\star}$, of the inner accretion disc and the central protostar,
\begin{eqnarray}\label{eq:Msink}
M\tsc{sink} = M\tsc{iad} + M_{\star},\\\label{eq:Lsink}
\bs{L}\tsc{sink} = \bs{L}\tsc{iad} + \bs{L}_{\star} \, .
\end{eqnarray}
Gas accreted by the sink particle is initially stored in the IAD. This gas may then be accreted onto the central protostar via two accretion channels, 
\begin{eqnarray}\label{eq:dmdt}
\frac{dM_{\star}}{dt}&=&\left.\frac{dM}{dt}\right|_{\textsc{bg}}\,+\,\left.\frac{dM}{dt}\right|_{\textsc{mri}} \, .
\end{eqnarray}
The background accretion rate, $\left.\frac{dM}{dt}\right|_{\textsc{bg}} = 10^{-7}  \,\Ms \, \mrm{yr}^{-1}$, allows for low but continuous accretion of gas onto the central protostar. The additional episodic accretion rate is much higher, on average $\left.\frac{dM}{dt}\right|_{\textsc{mri}} \simeq 5 \times 10^{-4}  \,\Ms \, \mrm{yr}^{-1}$, but only contributes during outburst events, which typically last a few tens of years. \cite{Stamatellos12a} assume that a combination of gravitational and magneto-rotational instabilities (MRI) acts as the main trigger for outbursts \citep{Zhu09, Zhu10}. In this way we obtain realistic accretion rates, similar to those observed in FU Orionis Type stars \citep{Bell94}. We use the episodic accretion rate for the following sub-grid models. Varying $\left.\frac{dM}{dt}\right|_{\textsc{mri}}$ has little effect on the outcome of the simulations \citep{Rohde18}.

%%%%%
\subsection{Stellar evolution model}\label{SEC:StellarEvolution}
%%%%%

Improving upon \cite{Rohde18} we implement the one-zone, stellar evolution model described in \cite{Offner09}, originally introduced by \cite{Nakano95}, and subsequently improved by \cite{Nakano00} and \cite{Tan04}. This sub-grid model describes the evolution of the stellar radius, $R_{\star}$, and luminosity, $L_{\star}$, due to the energy balance between accretion, gravitational contraction, nuclear burning, ionisation and radiation. The change in protostellar radius, $\dot{R}_{\star}$, is given by
\begin{eqnarray}
\begin{split}
    \dot{R}_{\star} =  \frac{2}{ M_{\star}}\frac{dM_{\star}}{dt} \left( 1 - \frac{1-f\tsc{k}}{a\tsc{g}(n) \beta } + \frac{1}{2} \frac{d \, \mrm{log} \, \beta}{d \, \mrm{log} \, M_{\star}}\right) R_{\star} \\ 
     - \frac{2}{a\tsc{g}(n) \beta} \left( \frac{R_{\star}}{\mrm{G} M_{\star}^2} \right) \left( L\tsc{int} + L\tsc{di} - L\tsc{db} \right) R_{\star} \, .
\end{split}    
\end{eqnarray} 
Here, $G$ is the gravitational constant, $f\tsc{k} = 0.5$ is the fraction of kinetic energy that is radiated away in the inner accretion disc, $a\tsc{g}(n)$ is the gravitational energy coefficient for a sphere with polytropic index  $n < 5$ \citep[][and references therein]{Nakano00}. $\beta$ is the ratio of gas pressure to total pressure (gas plus radiation) in the protostar. $L\tsc{int}$ is the internal luminosity, $L\tsc{di}$ is the power required to dissociate and ionise the accreted gas, and $L\tsc{db}$ is the power released by deuterium burning. 

This model follows the protostellar evolution through six distinct phases: (i) the initial `pre-collapse' phase; (ii) the `no burning' phase; (iii) the `core deuterium burning at fixed $\mrm{T}\tsc{c}$' phase; (iv) the `core deuterium burning at variable $\mrm{T}\tsc{c}$' phase; (v) the `shell deuterium burning' phase; and (vi) the `zero-age main sequence' phase \citep{Tout96}.

We follow the implementation described by \citet{Offner09} and also used by \citet{Murray18} and \citet{Cunningham18}. However, we use the mass of the sub-grid protostar, $M_{\star}$ (rather than the mass of the sink particle, $M\tsc{sink}$) and the episodic accretion rate from the IAD onto the protostar, $\frac{dM_{\star}}{dt}$ (rather than the sink particle's accretion rate, $\frac{dM\tsc{sink}}{dt}$; Section~\ref{SEC:EpisodicAccretion}). The accretion luminosity depends linearly on the accretion rate and is therefore highly variable due to the episodic nature of accretion onto the protostar.

%%%%%
\subsection{Radiative feedback}\label{SEC:RadFeedback}
%%%%%

Radiative feedback from young protostars can heat and stabilise their surrounding accretion discs, thus suppressing further disc fragmentation \citep{Jones18}. Theoretical studies have shown that this reduces the number of brown dwarfs and low-mass protostars formed \citep{Chabrier03, Offner09, Rice11, Guszejnov16, Guszejnov17}. However, continuous radiative feedback (i.e. neglecting episodic accretion effects) tends to suppress the formation of brown dwarfs and low-mass protostars too efficiently, resulting in a lower stellar multiplicity than observed  \citep{Stamatellos12a, Lomax14, Lomax15, Mercer17}. 

We make use of the episodic accretion model in \cite{Stamatellos12a} (Section~\ref{SEC:EpisodicAccretion}), in combination with the stellar evolution model in \cite{Offner09} (Section~\ref{SEC:StellarEvolution}) to compute the highly variable protostellar luminosities. These luminosities are taken into account by invoking a pseudo background radiation field with temperature, $T\tsc{bg}$. At general position $\bs{r}$, $\,T\tsc{bg}$ is given by
\begin{eqnarray}
    T^4\tsc{bg}(\bs{r}) = (10 \, \mrm{K})^4 + \sum_n \left( \frac{L_{\star, n}}{16 \pi \sigma\tsc{sb} |\bs{r} - \bs{r}_{\star, n} |^2}  \right)
\end{eqnarray}
\citep{Stamatellos07}. Here, $\bs{r}_{\star, n}$ and $L_{\star, n}$ are the position and luminosity of the $n^{\mathrm{th}}$ protostar. In the vicinity of a protostar $T\tsc{bg}$ decreases with distance $d$ from the protostar approximately as $d^{\,-1/2}$. This method will not capture accurately the radiative feedback from massive stars. However, we are interested here in the formation of low mass stars (our initial core mass is just 1 $\Ms$) and the model has been extensively tested in this regime \citep{Stamatellos12a, Lomax14, Lomax15, Mercer17, Rohde18}. 

%%%%%
\subsection{Outflow feedback}\label{SEC:OutflowFeedback}
%%%%%

We use the sub-grid episodic outflow model presented in \cite{Rohde18} with a few modifications. Here, we briefly outline the model and focus on the modifications. A more detailed description, including a parameter and resolution study, can be found in \cite{Rohde18}.

As in most sub-grid outflow models we assume that the mass ejection rate is a fixed fraction of the accretion rate,
\begin{eqnarray}
\left.\frac{dM}{dt}\right|_{\textsc{eject}}  = f\tsc{eject} \, \frac{dM_{\star}}{dt}  \, .
\end{eqnarray}
Here we adopt the default value $f\tsc{eject}=0.1$, based on observations and theoretical studies \citep[see][or the review by \citealt{Bally16}]{Croswell87, shu88, Pelletier92, Calvet93,  Hartmann95, Nisini18}. In contrast to most other sub-grid outflow models, we do not use the accretion rate onto the sink particle, $\frac{dM\tsc{sink}}{dt}$, but the episodic accretion rate onto the central star, $\frac{dM_{\star}}{dt}$ (Section~\ref{SEC:EpisodicAccretion}). This leads to the intermittent ejection of individual outflow bullets \citep{Rohde18}. To model the density and velocity distribution of the outflowing gas, we use the prescription for hydrodynamical outflows derived by \cite{Matzner99}. In this way we obtain a two component outflow, with a collimated high-velocity jet, and a low-velocity wide-angle disc wind.

For the outflow velocity we assume  
\begin{eqnarray}\label{eq.Vout}
\upsilon\tsc{out} = \left( \frac{\mrm{G} \, M_{\star}}{r\tsc{launch}} \right)^{1/2} \, P(\theta) \, ,
\end{eqnarray}
which is the Keplerian velocity at radius $r\tsc{launch}$, modulated with the angular distribution, $P(\theta)$, derived by \cite{Matzner99}. Here, $\theta$ is the angle at which the SPH particle is ejected relative to the spin axis of the accretion disc. In contrast to \cite{Rohde18} we do not adopt a fixed value for the launching radius, $r\tsc{launch}$. Instead we use a time-dependent radius depending on the stellar radius, $R_\star$, provided by the stellar evolution model,
\begin{eqnarray}\label{eq.Rlaunch}
r\tsc{launch} = 2 \, R_{\star} \, .
\end{eqnarray}
This gives us a more physically motivated outflow velocity, and avoids the need to invoke an arbitrary launching radius.

Outflows play a crucial role in removing angular momentum from the gas that is about to be accreted \citep{Hartmann89, Matt05}. Recent observations show that outflows are rotating and thus carry away angular momentum \citep{Launhardt09, Chen16, Lee17, Tabone17}. We incorporate rotating outflows by adding to the outward velocity, $\bs{\upsilon} \tsc{out}$, a rotational velocity component,
\begin{eqnarray}\label{eq.Vrot}
\bs{\upsilon} \tsc{rot}  = \bs{r} \times \bs{\omega} \, 
\end{eqnarray}
with
\begin{eqnarray}\label{eq.Omega}
\bs{\omega} = \frac{\ell\tsc{sph}}{m\tsc{sph} \, sin^2(\theta) \, r^2} \; \bs{\hat{\rm e}}_{\textsc{iad}} \, .
\end{eqnarray}
Here, $m\tsc{sph}$ is the mass of an SPH particle and $\hat{\rm e}_{\textsc{iad}} = \bs{L}\tsc{iad} / |\bs{L}\tsc{iad}|$ is the spin axis of the IAD. In contrast to \cite{Rohde18}, we calculate the angular momentum each ejected particle carries away, $\ell\tsc{part}$, from the break-up angular momentum of the protostar,
\begin{eqnarray}\label{eq.Lbreak}
L\tsc{breakup} = M_{\star} \, \sqrt{G \, M_{\star} \, R_{\star}}\,;
\end{eqnarray} 
this assumes that the protostar rotates at its break-up angular speed. Whenever angular momentum is accreted from the IAD onto the central protostar, we compute the excess angular momentum, 
\begin{eqnarray}\label{eq.Lpart}
\ell\tsc{sph} = \frac{ |\bs{L}_{\star}| - L\tsc{breakup}}{N\tsc{eject}}\,,
\end{eqnarray} 
allocate it to the ejected particles, and reduce $|\bs{L}_{\star}|$ to $L\tsc{breakup}$; $N\tsc{eject}$ is the number of ejected particles during this timestep.

%%%%%
\subsection{Simulation setup}\label{SEC:setup}
%%%%%

We have performed 88 simulations with different initial conditions or physical processes. All simulations start from a spherically symmetric, dense core with $M\tsc{core} = 1 \Ms$ embedded in a low density envelope at T = 10 K. The density profile follows the radial distribution of a Bonnor-Ebert sphere \citep[BES;][]{Bonnor56, Ebert57}. To obtain cores with $M\tsc{core} = 1 \, \Ms$, we first construct a critical BES, truncated at the critical dimensionless radius $\xi_0 = 6.5$. The central densities are chosen in such a way, that the masses of the BESs are $\Ms /3$, $\Ms /4$ and $\Ms /5$, corresponding to physical core radii of $r\tsc{core} = 0.017\,\pc,\;0.013\,\pc\;{\rm and}\;0.010\,\pc$, respectively. Then we increase the central densities by factors of $\fbes = 3,\;4\;{\rm or}\;5$, respectively, to $\rho\tsc{central} =  2.0\pow{-17} \, \gcm,\;4.8\pow{-17}\,\gcm\;{\rm or}\;9.4 \pow{-17}\,\gcm$ so that all the cores have $M\tsc{core} = 1 \, \Ms$. This makes the cores more and more super-critical with increasing  $\fbes$. Thus, cores with higher $\fbes$ are smaller, denser and have shorter free fall times, respectively, $t\tsc{ff} = 36.8\,{\rm kyr},\;24.6\,{\rm kyr}\;{\rm and}\;16.6\,{\rm kyr}$.

At $r\tsc{core}$, the radial density profile decreases smoothly but quickly (powerlaw with index $\gamma\!=\!-\,4$) to $\rho\tsc{env} = 10^{-23} \, \gcm$. The envelope then extends to $r\tsc{env} = 0.75 \, \pc$, which allows us to study the interaction of outflows with a low-density ambient medium. The total mass of the core plus envelope is $M\tsc{total} \sim 1.86 \, \Ms$; this mass varies by at most $0.2\%$ due to varying $r\tsc{core}$. 

As in \cite{Walch10}, we add an isotropic random Gaussian velocity field to the dense cores, in order to study the influence of turbulence on core collapse. The amplitudes follow a power-spectrum of the form
\begin{eqnarray}
P_{k} \propto k^{-4}\;\;\;\;\mathrm{with}\;\;\;k \in\left[\kmin,\,64\right] \, .
\end{eqnarray}
Due to the steep power-spectrum, most of the turbulent energy is associated with the smallest wavenumber, $\kmin$. We stipulate $\kmin = 1,\;2\;{\rm or}\;3$, with $\frac{2 \pi}{\kmin} = 1$ corresponding to the core diameter. In this way we change the velocity field from large-scale motions ($\kmin = 1$) with high net angular momentum, to small-scale turbulence ($\kmin = 3$) with low net angular momentum \citep{Walch12}. We vary the strength of the turbulence by adjusting the virial ratio
\begin{eqnarray}
\avir = \frac{2\,(E\tsc{turb} + E\tsc{therm})}{\left|E\tsc{grav}\right|} \, .
\end{eqnarray}

We perform simulations for all combinations of $\avir = 0.5, \;1.0, \;2.0 \;{\rm and} \;3.0$, $\kmin = 1,\;2\;{\rm and}\;3$, and $\rc = 0.010\,\pc,\;0.013\,\pc\;{\rm and}\;0.017\,\pc$. In addition, we perform runs with $\avir = 1.0$, $\kmin = 1$, and  $\rc = 0.013$ pc for eight different turbulent seeds. To study the influence of outflow feedback on the SFE we produce a comparison run without outflow feedback for each setup. This adds up to 88 simulations in total (see Table \ref{Table:Sims}). The mass resolution is $400\,000\,$ SPH particles per $\Ms$, resulting in a total number of $N\tsc{total} \sim  740\,000$ SPH particles.

\begin{table}
\caption{Parameter summary for all the simulations performed. Reading from left to right the columns give the run number, the run name, the turbulent random seed ($\chi$), the virial ratio ($\avir$), the smallest turbulent wavenumber ($\kmin$), and the core radius ($\rc/{\rm pc}$). Each combination of parameters is simulated once with, and once without, outflow feedback. The simulations with outflow feedback have odd IDs and their run names end with ${\rm x}=1$. The simulations without outflow feedback have even IDs and their run names end with ${\rm x}=0$.} 
\hspace{-0.5cm}
\begin{center}\begin{tabular}{clcccc}
\hline 
\# & Run & $\chi$ & $\avir$ & $\kmin$ & $\rc$ \\
\hline 
1/2   & \sn{1}{0.5}{1}{0.017}{x}   & 5 & 0.5 & 1 & 0.017 \\
3/4   & \sn{1}{1.0}{1}{0.017}{x}   & 5 & 1.0 & 1 & 0.017 \\
5/6   & \sn{1}{2.0}{1}{0.017}{x}   & 5 & 2.0 & 1 & 0.017 \\
7/8   & \sn{1}{3.0}{1}{0.017}{x}   & 5 & 3.0 & 1 & 0.017 \\
9/10  & \sn{1}{0.5}{2}{0.017}{x}   & 5 & 0.5 & 2 & 0.017 \\
11/12 & \sn{1}{1.0}{2}{0.017}{x}   & 5 & 1.0 & 2 & 0.017 \\
13/14 & \sn{1}{2.0}{2}{0.017}{x}   & 5 & 2.0 & 2 & 0.017 \\
15/16 & \sn{1}{3.0}{2}{0.017}{x}   & 5 & 3.0 & 2 & 0.017 \\
17/18 & \sn{1}{0.5}{3}{0.017}{x}   & 5 & 0.5 & 3 & 0.017 \\
19/20 & \sn{1}{1.0}{3}{0.017}{x}   & 5 & 1.0 & 3 & 0.017 \\
21/22 & \sn{1}{2.0}{3}{0.017}{x}   & 5 & 2.0 & 3 & 0.017 \\
23/24 & \sn{1}{3.0}{3}{0.017}{x}   & 5 & 3.0 & 3 & 0.017 \\
25/26 & \sn{1}{0.5}{1}{0.013}{x}   & 5 & 0.5 & 1 & 0.013 \\
27/28 & \sn{1}{1.0}{1}{0.013}{x}   & 5 & 1.0 & 1 & 0.013 \\
29/30 & \sn{1}{2.0}{1}{0.013}{x}   & 5 & 2.0 & 1 & 0.013 \\
31/32 & \sn{1}{3.0}{1}{0.013}{x}   & 5 & 3.0 & 1 & 0.013 \\
33/34 & \sn{1}{0.5}{2}{0.013}{x}   & 5 & 0.5 & 2 & 0.013 \\
35/36 & \sn{1}{1.0}{2}{0.013}{x}   & 5 & 1.0 & 2 & 0.013 \\
37/38 & \sn{1}{2.0}{2}{0.013}{x}   & 5 & 2.0 & 2 & 0.013 \\
39/40 & \sn{1}{3.0}{2}{0.013}{x}   & 5 & 3.0 & 2 & 0.013 \\
41/42 & \sn{1}{0.5}{3}{0.013}{x}   & 5 & 0.5 & 3 & 0.013 \\
43/44 & \sn{1}{1.0}{3}{0.013}{x}   & 5 & 1.0 & 3 & 0.013 \\
45/46 & \sn{1}{2.0}{3}{0.013}{x}   & 5 & 2.0 & 3 & 0.013 \\
47/48 & \sn{1}{3.0}{3}{0.013}{x}   & 5 & 3.0 & 3 & 0.013 \\
49/50 & \sn{1}{0.5}{1}{0.010}{x}   & 5 & 0.5 & 1 & 0.010 \\
51/52 & \sn{1}{1.0}{1}{0.010}{x}   & 5 & 1.0 & 1 & 0.010 \\
53/54 & \sn{1}{2.0}{1}{0.010}{x}   & 5 & 2.0 & 1 & 0.010 \\
55/56 & \sn{1}{3.0}{1}{0.010}{x}   & 5 & 3.0 & 1 & 0.010 \\
57/58 & \sn{1}{0.5}{2}{0.010}{x}   & 5 & 0.5 & 2 & 0.010 \\
59/60 & \sn{1}{1.0}{2}{0.010}{x}   & 5 & 1.0 & 2 & 0.010 \\
61/62 & \sn{1}{2.0}{2}{0.010}{x}   & 5 & 2.0 & 2 & 0.010 \\
63/64 & \sn{1}{3.0}{2}{0.010}{x}   & 5 & 3.0 & 2 & 0.010 \\
65/66 & \sn{1}{0.5}{3}{0.010}{x}   & 5 & 0.5 & 3 & 0.010 \\
67/68 & \sn{1}{1.0}{3}{0.010}{x}   & 5 & 1.0 & 3 & 0.010 \\
69/70 & \sn{1}{2.0}{3}{0.010}{x}   & 5 & 2.0 & 3 & 0.010 \\
71/72 & \sn{1}{3.0}{3}{0.010}{x}   & 5 & 3.0 & 3 & 0.010 \\
73/74 & \sn{2}{1.0}{1}{0.013}{x}   & 0 & 1.0 & 1 & 0.013 \\
75/76 & \sn{3}{1.0}{1}{0.013}{x}   & 1 & 1.0 & 1 & 0.013 \\
77/78 & \sn{4}{1.0}{1}{0.013}{x}   & 2 & 1.0 & 1 & 0.013 \\
79/80 & \sn{5}{1.0}{1}{0.013}{x}   & 3 & 1.0 & 1 & 0.013 \\
81/82 & \sn{6}{1.0}{1}{0.013}{x}   & 4 & 1.0 & 1 & 0.013 \\
83/84 & \sn{7}{1.0}{1}{0.013}{x}   & 6 & 1.0 & 1 & 0.013 \\
85/86 & \sn{8}{1.0}{1}{0.013}{x}   & 7 & 1.0 & 1 & 0.013 \\
87/88 & \sn{9}{1.0}{1}{0.013}{x}   & 8 & 1.0 & 1 & 0.013 \\
\hline
\end{tabular}\end{center} 
\label{Table:Sims}
\end{table}

%%%%%
\section{Results}\label{SEC:Results}
%%%%%

Due to their different initial conditions, some simulations form stars faster than others. To carry out objective comparisons between the simulations, we make them at times $\,t_\tau$ where $t_\tau=t_0+\tau\tff$; here $t_0$ is the time at which the first sink forms, $\tff = \pi \, (\rc^3/(8GM_{\textsc{core}})^{1/2}$ is the core's freefall time, and we use $\tau =0.5,\;1.5\;{\rm and}\;5.0$. All simulations are terminated at $t_5\sim 200\pm 50\,{\rm kyr}$. The ensemble of simulations is divided into those with outflow feedback (the OF-sample, with odd IDs and run names ending in `O-1') and those without outflow feedback (RF-sample, with even IDs and run names ending in `O-0'). Both samples contain 44 simulations (Table \ref{Table:Sims}). The OF-sample forms \nolinebreak[4]{$N_{\star{\rm O}-1} = 132$} stars in total, whereas the RF-sample forms $N_{\star{\rm O}-0} = 163$. All statistical tests use a significance threshold of $p<1\%$.

%%%%%
\subsection{Stellar diversity}\label{SEC:VR}
%%%%%

\begin{figure*}
\centering
\includegraphics[width=0.80\textwidth]{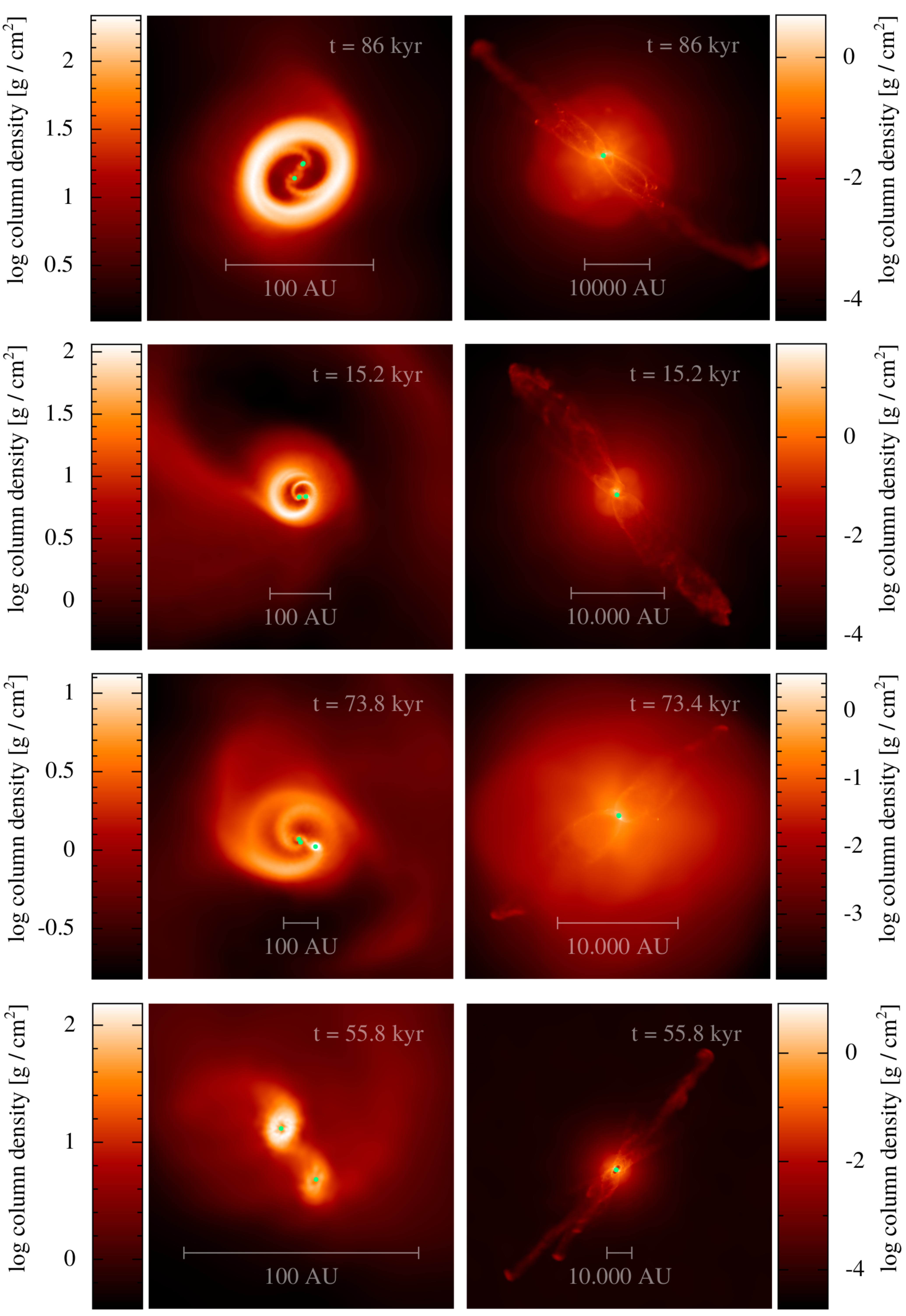}
\caption{\label{fig:ColDens} 
Column density plots of four representative simulations with outflow feedbackat $t_{0.5}\equiv t_0+0.5\,\tff$. The left column shows the multiple systems and accretion discs in the central regions. The right column shows the same simulations, but zoomed out to reveal their outflows. The green dots represent sink particles. The simulations are from top to bottom \sn{5}{2.0}{3}{0.017}{1}, \sn{5}{0.5}{1}{0.010}{1}, \sn{5}{3.0}{1}{0.013}{1} and \sn{5}{1.0}{3}{0.017}{1}. Note that the scale, the colour bar and the viewing-angle are different for each panel.}
\end{figure*}

The ensemble of simulations produces a wide variety of stellar configurations: single stars and multiple systems; circumstellar and/or circumbinary discs; aligned and misaligned outflows. Fig.~\ref{fig:ColDens} illustrates four representative runs with outflow feedback, all at $t_{0.5}$. The green markers represent sink particles. The left column shows the central regions around the sink particles and their accretion discs. The right column shows the  outflows on larger scales.

The simulation on the top row of Fig.~\ref{fig:ColDens} (\sn{5}{2.0}{3}{0.017}{1}) forms a wide binary system with a circumbinary disc. A third star forms in the circumbinary disc via disc-fragmentation. The binary system becomes an hierarchical triple system when the third star spirals inwards. The outflows from all three stars are well aligned and produce a broad outflow cavity. 

The simulation on the second row of Fig.~\ref{fig:ColDens} (\sn{5}{0.5}{1}{0.010}{1}) also forms three stars that end up in a stable triple system. Two of these stars belong to a close binary system with a circumbinary disc, while the third star has its own circumstellar disc. These two systems are surrounded by a larger accretion disc, and material from this larger disc streams inwards along a spiral structure and onto the two smaller discs.

The simulation on the third row of Fig.~\ref{fig:ColDens} (\sn{5}{3.0}{1}{0.013}{1})  forms four stars in total. Initially these stars are in an hierarchical quadruple system (a close binary, a third star orbiting further out, and a fourth star orbiting even further out). This fourth star has the largest accretion disc, and there are spiral accretion flows feeding material inwards from larger scales and onto the accretion discs. Later on only the close binary remains bound.

The simulation on the bottom row of Fig.~\ref{fig:ColDens} (\sn{5}{1.0}{3}{0.017}{1}) forms two stars in a binary system. Both stars have their own circumstellar accretion discs, with a bridge in between. At this stage the outflows from the stars point in slightly different directions, but later on they align. 

%%%%%

%%%%%
\subsection{Overview of stellar masses and multiplicities}\label{SEC:StarFormation}
%%%%%

\begin{figure*}
    \centering
    \includegraphics[width=1.0\textwidth]{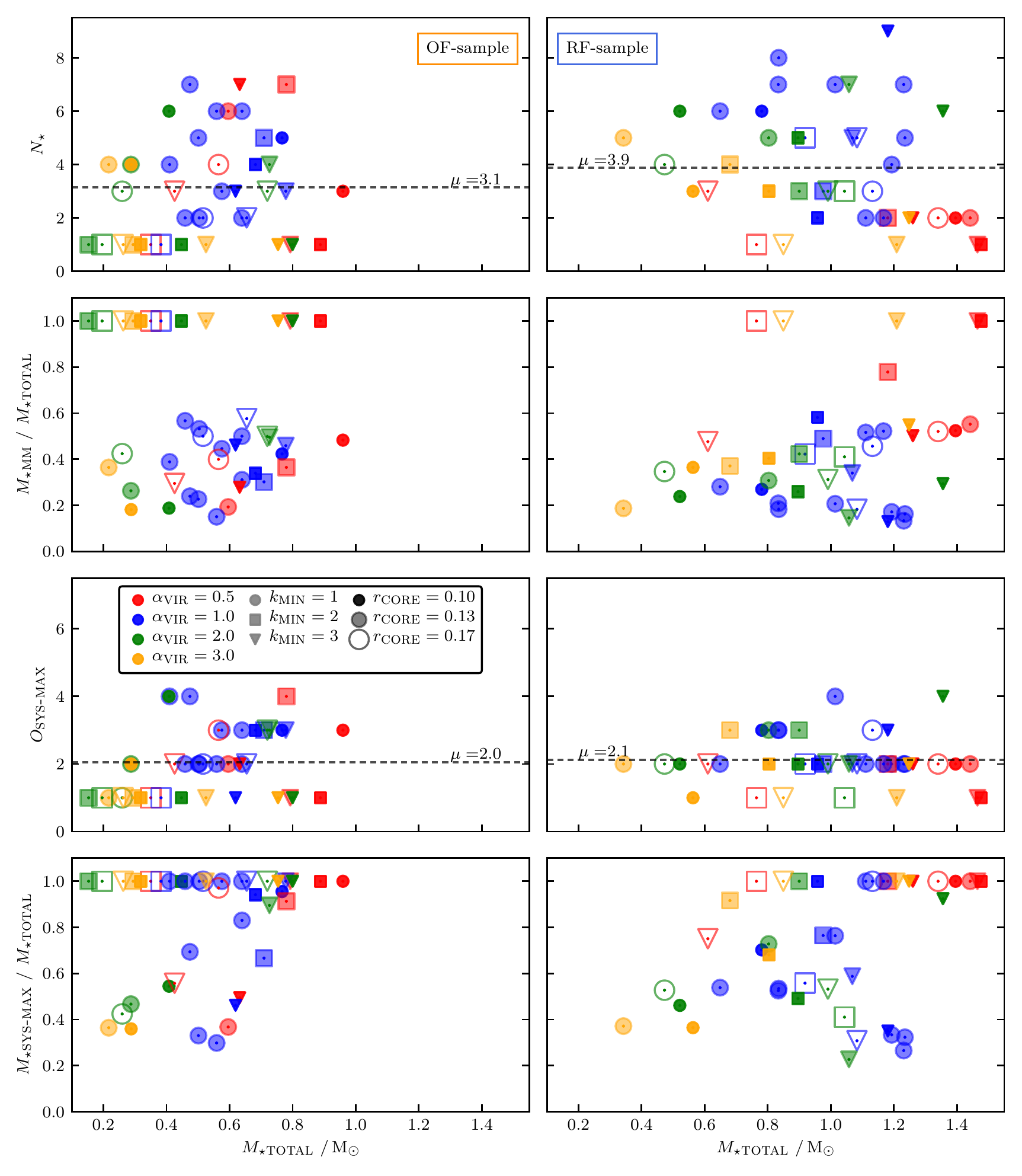}
    \caption{\label{fig:SF} Scatter plots showing the properties of the stars formed in a core against total mass of stars, for all the simulations, at $t _5\equiv t_0+5\,\tff$. The left column shows the OF-sample, and the right column the RF-sample. The size of the symbol encodes the core's initial radius, $\rc$; the colour of the symbol encodes the core's initial virial parameter, $\avir$; and the shape of the symbol encodes the wavenumber of the largest initial turbulent mode in the core (see key on third panel down, lefthand side). The top row shows the total number of stars, $N_\star$, and the dashed lines indicate the mean values, $\bar{N}_\star\sns{O}{1} = 3.14 \pm 1.95$ and $\bar{N}_\star\sns{O}{0} = 3.88 \pm 2.12$. The second row shows the mass of the most massive star, as a fraction of the total mass of stars, $M_{\star \textsc{mm}}/M_{\star \textsc{total}}$. The third row shows the order of the highest-order system, and the dashed lines indicate the mean values, $\bar{O}\tsc{sys}\sns{O}{1} = 2.05 \pm 1.00$ and $\bar{O}\tsc{sys}\sns{O}{0} = 2.12 \pm 0.73$. The bottom row shows the mass of the highest-order system, as a fraction of the total mass of stars, $M_{\star \textsc{sys-max}}/M_{\star \textsc{total}}$.}
\end{figure*}

Fig.~\ref{fig:SF} shows, as a function of the total mass in stars, $M_{\star \textsc{total}}$ at $t_5$, the number of stars formed in a core, $N_{\star}$ (top row), the mass of the most massive star, $M_{\star \textsc{mm}}$ (second row), the order of the highest order system, $O_{\textsc{sys-max}}$ (third row), and the total stellar mass of this highest-order system, $M_{\star \textsc{sys-max}}$ (bottom row). The left column shows the results for the OF-sample, and the right column shows them for the RF-sample. The star formation efficiency is $\mbox{\sc sfe}=M_{\star\textsc{total}}/M\tsc{core}$ at $t_5$. Since the core can accrete matter from its surroundings and convert this matter into stars, {\sc sfe} can exceed unity.

The top row of Fig.~\ref{fig:SF} indicates that there is no significant correlation between $N_{\star}$ and $M_{\star \textsc{total}}$, for either sample.  A Kendall Rank Correlation (KRC) test confirms this, with $p\sim 30\%$ for both samples. On average, the RF-sample forms more stars per core, $\bar{N}_{\star}$\sns{O}{0}$=3.88\pm\,2.12$ than the OF-sample, $\bar{N}_{\star}$\sns{O}{1}$=3.14\pm\,1.95$. The theoretical model of \cite{Holman13} predicts  a slightly higher number, $N_{\star} = 4.1 \pm 0.4$. The core that forms the highest number of stars, $N_{\star} = 9$, is \sn{5}{1.0}{3}{0.010}{0} in the RF-sample. The OF-sample contains 13 simulations which form only a single star, as compared with only 5 in the RF-sample.

%%%%%   I would prefer SD to error   %%%%%   <<<<<

The second row of Fig.~\ref{fig:SF} shows that the ratio of the mass of the most massive star to the total stellar mass, $M_{\star \textsc{mm}}/M_{\star \textsc{total}}$, is between $\sim 0.2$ and $\sim 0.6$. For the OF-sample, this ratio shows no correlation with the total stellar mass, $M_{\star \textsc{total}}$. For the RF-sample, the ratio shows a slight tendency to increase with increasing $M_{\star \textsc{total}}$, but with a large scatter.

The third row of Fig.~\ref{fig:SF} shows that the order of the highest-order system formed in each core, $O_{\textsc{sys-max}}$, is not significantly correlated with $M_{\star \textsc{total}}$.\footnote{The highest-order system formed in a core is not necessarily a higher-order multiple (HOM), it could be a single or a binary.} Multiple systems are identified and characterised using the method proposed by \cite{Lomax15}, which iteratively pairs up stars and multiples in an hierarchical order, taking into account their mutual gravitational and kinetic energies, their eccentricity, and whether the pair is tidally bound. Many single stars are ejected by dynamical interactions with multiples. We discuss these ejected stars in Section~\ref{SEC:Multiplicity}. 

Strictly speaking, only binary systems are truly stable, in the sense that they can survive indefinitely, in isolation. However, HOMs can survive for a very long time if they are arranged hierarchically. Consequently, some of them will survive long after the dispersal of the birth core, but many will end up as binaries, and some will dissolve completely into singles (e.g. run \rm{\sn{5}{1.0}{3}{0.010}{1}}). In general, the larger the number of stars, the larger the number of ejected singles. For example, run \sn{5}{1.0}{3}{0.010}{0} forms 9 stars, but ejects 6 of them and ends up as an hierarchical triple system. In all simulations that form only two stars, these two always end up in a binary.

One very striking difference between  the OF- and RF-samples is the fractions of single ($S_1$), binary ($B_2$), triple ($T_3$) and quadruple ($Q_4$) systems formed. For the OF-sample, there is a monotonic decrease with increasing order, viz. $(S_1:B_2:T_3:Q_4) = (0.38:0.29:0.24:0.10)$. In contrast, the RF-sample mainly forms binary systems, $(S_1:B_2:T_3:Q_4) = (0.17:0.60:0.19:0.05)$. The fraction of triple and quadruple systems is slightly higher for the OF-sample. However, due to the high fraction of binaries in the RF-sample, the mean orders of the largest systems are very similar: $O_{\textsc{sys-max}}=2.0$ for the OF-sample, and $O_{\textsc{sys-max}}=2.1$ for the RF-sample.

The bottom row of Fig.~\ref{fig:SF} shows the ratio of the mass in the highest-order system to the total stellar mass, $M_{\star \textsc{sys-max}}/M_{\star \textsc{total}}$. The OF-sample has significantly more simulations with $M_{\star \textsc{sys-max}}/M_{\star \textsc{total}}=1.0$, because many more simulations form just a single star. Setting aside the systems with a ratio close to one, $M_{\star \textsc{sys-max}}/M_{\star \textsc{total}}$ tends to increase with increasing $M_{\star \textsc{total}}$, up to $M_{\star \textsc{total}}\sim 0.8\Ms$, for both samples, albeit with large scatter. Above $M_{\star \textsc{total}}\sim 0.8\Ms$, there are no multiple systems in the OF-sample, but for the RF-sample $M_{\star \textsc{sys-max}}/M_{\star \textsc{total}}$ then tends to decrease with increasing $M_{\star \textsc{total}}$; this is because these simulations produce large numbers of stars and only a few of them end up in the highest-order system.

%%%%%
\subsection{Influence of initial core properties}\label{SEC:CoreProp}
%%%%%

The initial conditions for the simulated cores are characterised by four parameters: the core radius, which takes values $\rc = 0.010\,{\rm pc},\;0.013\,{\rm pc}\;{\rm and}\;0.017\,{\rm pc}$; the virial ratio, which takes values $\avir = 0.5,\;1.0,\;2.0\;{\rm and}\; 3.0$; the minimum wavenumber for the imposed turbulent modes, which takes values $\kmin = 1,\;2\;{\rm and}\;3$; and the seed for the random turbulent modes excited, which takes the same value $\chi =1$ for all combinations of $(\rc,\avir,\kmin)$ except for $(\rc,\avir,\kmin)=(0.013\,{\rm pc},1.0,1)$, for which we perform runs with $\chi =1,\;2,\;3,\;4,\;5,\;6,\;7\;{\rm and}\;8$.

\begin{figure*}
    \centering
    \includegraphics[width=1.0\textwidth]{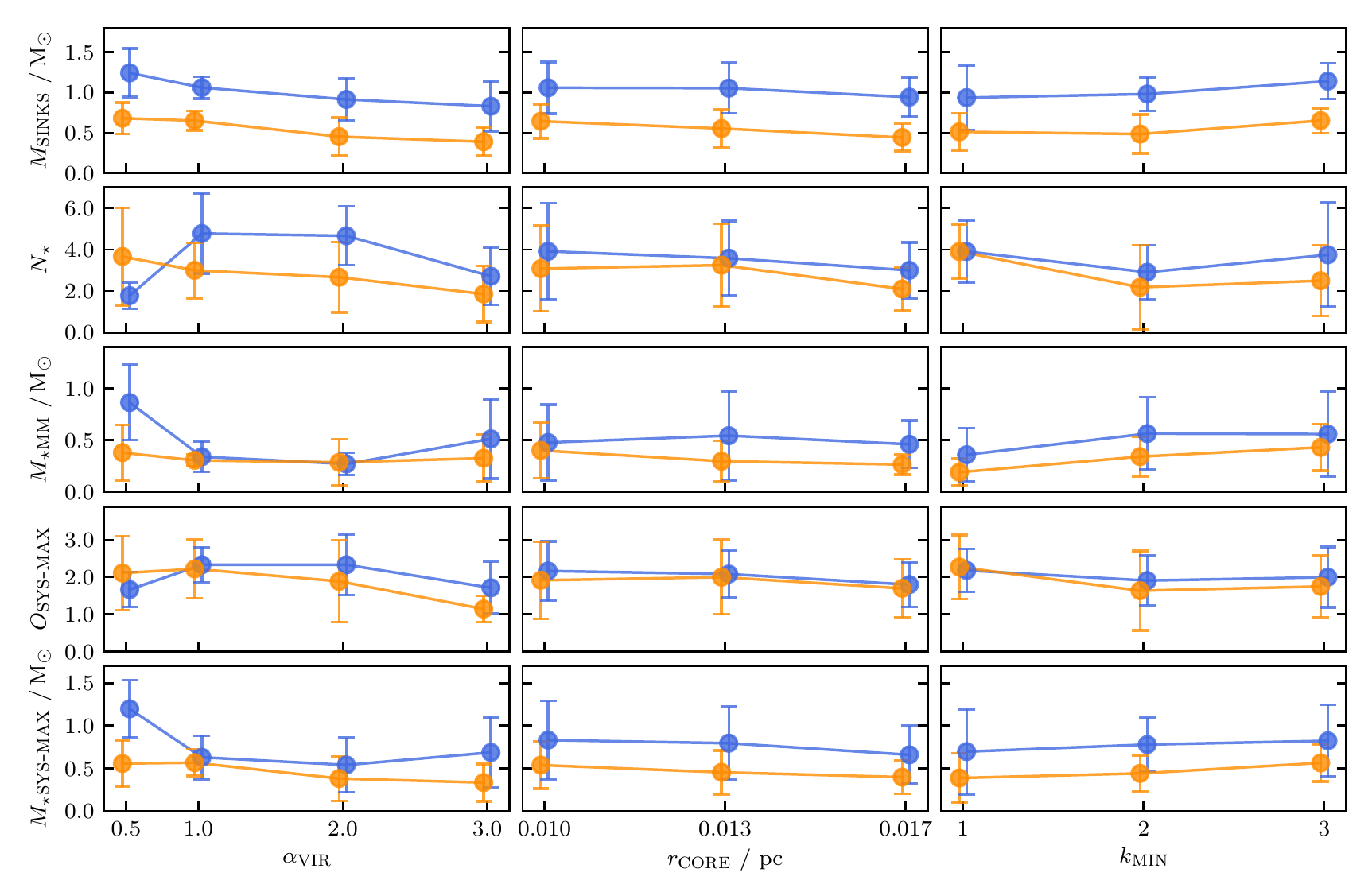}
    \caption{\label{fig:SF-Mean} Mean values for the stellar parameters plotted on Fig.~\ref{fig:SF}, as a function of the parameters defining the initial conditions of the birth core. Simulations from the OF-sample are plotted in orange, and those from the RF-sample are plotted in blue. The left column shows the means for subsets with the same $\avir = 0.5,\;1.0,\;2.0\;{\rm or}\; 3.0$ (i.e. averaged over all values of $\rc$ and $\kmin$). The middle column shows the means for subsets with the same $\rc = 0.010\,{\rm pc},\;0.013\,{\rm pc}\;{\rm or}\;0.017\,{\rm pc}$ (i.e. averaged over all values of $\avir$ and $\kmin$). The right column shows the means for subsets with the same $\kmin = 1,\;2\;{\rm or}\;3$ (i.e. averaged over all values of $\avir$ and $\rc$). The top row shows the mean total stellar mass, $\bar{M}_{\star \textsc{total}}$; the second row shows the mean number of stars, $\bar{N}_\star$; the third row shows the mean mass of the most massive star, $\bar{M}_{\star \textsc{mm}}$; the fourth and fifth rows show the mean order, $\bar{O}\tsc{sys}$, and the mean mass, $\bar{M}_{\star \textsc{sys-max}}$, of the highest-order system. The initial conditions of the birth core appear to have very limited influence on the properties of the stars formed.}
\end{figure*}

In order to explore how these parameters influence the masses and multiplicities of the stars formed in a core, we compute average values at $t_{5}\!\equiv\!t_0\!+\!5t\tsc{ff}$ for (i) the total stellar mass, $M_{\star \textsc{total}}$, (ii) the number of stars, $N_\star$, (iii) the mass of the most massive star, $M_{\star \textsc{mm}}$, (iv) the order of the highest-order system, $O\tsc{sys}$, and (v) the mass of the highest order system, $M_{\star \textsc{sys-max}}$, for all the simulations with a given radius $\rc$ but different values of $\avir$ and $\kmin$ -- and similarly for all the simulations with a given virial parameter $\avir$ but different $\rc$ and $\kmin$, and all the simulations with a given minimum turbulent wavenumber $\kmin$  but different $\rc$ and $\avir$. The results are presented on Fig.~\ref{fig:SF-Mean}, where the results for simulations from the OF-sample are in orange, and those from the RF-sample are in blue.

To quantify the results presented in Fig.~\ref{fig:SF-Mean} we evaluate the dependence of these mean stellar parameters ($\bar{M}_{\star \textsc{total}},\bar{N}_\star,\bar{M}_{\star \textsc{mm}},\bar{O}\tsc{sys},\bar{M}_{\star \textsc{sys-max}}$) on the initial condition parameters ($\avir,\rc,\kmin$) by computing the Kendall Rank Correlation statistics, $\tau$ and $p$ (see Table \ref{Table:CoreProp}); $\tau$ gives the degree of correlation (or anti-correlation, if negative). In addition, we evaluate whether the OF- and RF-samples are drawn from the same underlying distribution, by computing the non-parametric Kolmogorov--Smirnov (KS) statistics, $d$ and $p$; $d$ measures the difference between the two distributions. In both cases, $p$ gives the probability of obtaining the evaluated correlation ($\tau$) or difference ($d$) assuming the null hypothesis (i.e. that both samples are drawn from the same underlying distribution).

Fig.~\ref{fig:SF-Mean} and Table \ref{Table:CoreProp} demonstrate clearly that varying the initial conditions -- at least in the range we have studied -- has little influence on the properties of the stars formed. The correlations between stellar parameters and initial condition parameters are at best weak ($|\tau| \le 40$), and in most cases they are not significant, so we only discuss those for which $p<1\%$. For both samples (OF and RF), $\bar{M}_{\star \textsc{total}}$ decreases with increasing $\avir$ (because the cores have more support and collapse more slowly); for the OF-sample, $\bar{M}_{\star \textsc{total}}$ also decreases with increasing $\rc$ (firstly because the cores collapse more slowly, and secondly because the outflow feedback acts on more rarefied gas and is therefore more effective). For the OF-sample, $\bar{N}_\star$ decreases with increasing $\avir$ (because the cores have more support and are therefore more easily dispersed by outflow feedback). $\bar{M}_{\star \textsc{mm}}$ increases with increasing $\kmin$ for the OF-sample (because the turbulence is concentrated on small scales which dissipate more rapidly), and with decreasing $\avir$ for the RF-sample (because the cores have less turbulent support and therefore their collapse is more focussed). $\bar{M}_{\star \textsc{sys-max}}$ increases with decreasing $\avir$ for both samples (again, because the cores have less turbulent support and therefore their collapse is more focussed).

The one exception to these correlations, anti-correlations and insignificant correlations is the $\avir =0.5$ RF subset, which bucks most of the trends seen in the other subsets. The very low level of core support ($\avir =0.5$) and the lack of outflow feedback (RF) result in a rather focussed infall onto the centre of the core, and consequently the formation of either a massive single star, or a massive binary (usually with approximately equal-mass components).

The second double column of Table \ref{Table:CoreProp} demonstrates that the OF- and RF-samples are statistically distinct. In particular, $\bar{M}_{\star \textsc{total}}$ is almost twice as large for the RF-sample as for the OF-sample (see Fig.~\ref{fig:SF-Mean}, top panel). $\bar{M}_{\star \textsc{mm}}$ and $\bar{M}_{\star \textsc{sys-max}}$ are also larger for the RF-sample than the OF-sample, by $\sim 50\%$ (see Fig.~\ref{fig:SF-Mean}, third and bottom panels). These differences are mainly due to the fact that in the OF simulations the outflow feedback disperses the gas surrounding the core, but in the RF simulations the surrounding gas falls onto the core and replenishes its mass. $\bar{N}_\star$ and $\bar{O}\tsc{sys}$ are indistinguishable between the two samples.

%%%%%
%%%%%

%%%%%
\begin{table*}
\caption{Non-parametric measures of the correlations, and their statistical significances, for the data presented in Fig.~\ref{fig:SF-Mean}. The left double-column gives the stellar parameters considered, in the same order as in Fig.~\ref{fig:SF-Mean}, and the sample used (outflow OF or reference RF). The second double-column gives the Kolmogorov--Smirnov statistics, $d$ and $p$, which reflect the likelihood that the two samples are drawn from the same distribution. The last three double-columns give the Kendall Rank Correlation (KRC) statistics, $\tau$ and $p$, which reflect the likelihood that the stellar parameters are correlated with -- respectively -- $\avir$, $\rc$, and $\kmin$; the KRC statistics are evaluated separately for the OF- and RF-samples. Correlations that satisfy our significance threshold of $p<1\%$ are highlighted.}
\label{Table:CoreProp}
\begin{tabular}{llC{0.5cm}rrC{0.7cm}rrC{0.7cm}rrC{0.7cm}rr}
\hline
\multicolumn{2}{c}{\multirow{2}{*}{subset}} && \multicolumn{2}{c}{KS-test} && \multicolumn{2}{c}{$\alpha_{\textsc{vir}}$} && \multicolumn{2}{c}{$r_{\textsc{core}}$} && \multicolumn{2}{c}{$k_{\textsc{min}}$} \\ 
\multicolumn{2}{c}{} && $d$ & $p$ [\%] && $\tau$ & $p$ [\%] && $\tau$ & $p$ [\%] && $\tau$ & $p$ [\%] \\ 
\hline                          
\hline 
\multirow{2}{*}{$M_{\star \textsc{total}}$} & OF && \multirow{2}{*}{$\bs{0.68}$} & \multirow{2}{*}{$\bs{<0.01}$} && $\bs{-0.40}$ & $\bs{0.06}$  &&  $\bs{-0.34}$ &  $\bs{0.45}$  &&  0.20        &  9.94  \\
                                            & RF &&                       &                                      && $\bs{-0.42}$ & $\bs{0.03}$  && -0.21         & 7.71          &&  0.18        & 12.39  \\
\hline                                                          
\multirow{2}{*}{$N_{\star}$}                & OF && \multirow{2}{*}{0.18} & \multirow{2}{*}{42.25}               && $\bs{-0.33}$ & $\bs{0.73}$  && -0.20         & 12.21         && -0.29        &  2.01  \\
                                            & RF &&                       &                                      &&  0.04        & 71.56        && -0.17         & 17.47         && -0.16        & 20.45  \\
\hline              
\multirow{2}{*}{$M_{\star \textsc{mm}}$}    & OF && \multirow{2}{*}{0.31} & \multirow{2}{*}{ 1.76}               && -0.15        & 18.89        && -0.17         & 14.37         &&  $\bs{0.39}$ &  $\bs{0.11}$  \\
                                            & RF &&                       &                                      && $\bs{-0.36}$ &  $\bs{0.16}$ && -0.04         & 69.44         &&  0.19        & 10.40  \\
\hline              
\multirow{2}{*}{$O_{\textsc{sys-max}}$}     & OF && \multirow{2}{*}{0.18} & \multirow{2}{*}{42.25}               && -0.30        &  1.69        && -0.13         & 31.62         && -0.30        &  2.27  \\
                                            & RF &&                       &                                      && -0.07        & 55.24        && -0.25         &  5.63         && -0.19        & 14.98  \\
\hline              
\multirow{2}{*}{$M_{\textsc{sys-max}}$}     & OF && \multirow{2}{*}{$\bs{0.36}$} & \multirow{2}{*}{$\bs{0.41}$}  && $\bs{-0.30}$ & $\bs{0.10}$  && -0.23         &  4.95         &&  0.22        &  7.18  \\ 
                                            & RF &&                       &                                      && $\bs{-0.41}$ & $\bs{0.01}$  &&- 0.19         & 11.61         &&  0.12        & 30.00  \\
\hline
\end{tabular}
\end{table*}
%%%%%

%%%%%
\subsection{Influence of turbulent seeds}\label{SEC:Seed}
%%%%%

To make sure our results are not dominated by the particular choice of the turbulent velocity field for the fiducial runs, we perform eight additional runs with different random turbulent seeds (Table \ref{Table:Sims}, runs with number 73-88), with and without outflow feedback. These runs have otherwise the same initial condition values as our fiducial runs \sn{5}{1}{1}{4}{x}. In Fig \ref{fig:SF} the subset of these runs ($\chi$-subset) are represented by the middle-sized, blue shaded circles. The spread of the ($\chi$-subset) is comparable to the spread of the runs with varying initial conditions (IC-subset). Remarkable is that the $\chi$-subset contains no run that forms only a single star.  

Table \ref{Table:TurbSeed} gives the mean and standard deviation of the full sample (OF and RF), the $\chi$-subset and the IC-subset for all quantities presented in Fig. \ref{fig:SF} and \ref{fig:SF-Mean}. The mean values and their spread are comparable for both samples. Two differences, however not statistically significant, are that the $\chi$-subset (a) forms on average slightly more stars and (b) has slightly lower masses of the most massive star due to the absence of runs forming a single star. The last two columns of Table \ref{Table:TurbSeed} give the Kolmogorov–Smirnov statistics, $d$ and $p$, reflecting the difference between the $\chi$- and IC-subset and the probability of finding these results assuming the null hypothesis is true. We do not find a statistical difference between the two subsets, and we are unable to reject the null hypothesis that both subsets have the same underlying distribution. However, we caution that this is not a proof that the distributions are the same.

The similarity between the $\chi$-subset and the IC-subset makes us confident that our results are not dominated by the choice of the random seeds. On the other hand, this finding supports our result from Section \ref{SEC:CoreProp}. Since the influence of the varying initial conditions on the outcome of the simulation is not higher than the influence due to different turbulent seeds, the core properties play at most a limited role in the outcome of the simulation.

%%%%%
\begin{table*}
\caption{Mean and standard deviation for the quantities presented in Fig. \ref{fig:SF} for the subsets with varying turbulent seeds and initial conditions (IC). The first column gives the quantities presented in Fig \ref{fig:SF}, the second column the feedback mechanism. The third, fourth and fifth column give the mean and standard deviation for the full OF- and RF- sample (full-sample), the reduced subset with varying turbulent seeds ($\chi$-subset) and the reduced subset with varying initial condition parameters (IC-subset). The sixth and seventh column give the Kolmogorov--Smirnov statistics, $d$ and $p$, which reflect the likelihood that the $\chi$- and IC-sample are drawn from the same distribution.}
\label{Table:TurbSeed}
\begin{tabular}{lcccccr}
\hline
quantity                                   & feedback & full-sample     &  $\chi$-subset     & IC-subset       &  $d$   &  $p [\%]$   \\
\hline
\multirow{2}{*}{$\bar{M}_{\star \textsc{total}}$}& OF & $0.54 \pm 0.20$ & $0.53 \pm  0.08$ & $0.54 \pm 0.22$ & 0.38 & 19.04 \\
                                                 & RF & $1.00 \pm 0.28$ & $1.03 \pm  0.20$ & $1.00 \pm 0.30$ & 0.24 & 74.01 \\
\hline
\multirow{2}{*}{$\bar{N}_{\star}$}               & OF & $3.14 \pm 1.95$ & $4.11 \pm  1.85$ & $2.85 \pm 1.87$ & 0.38 & 19.04 \\
                                                 & RF & $3.88 \pm 2.12$ & $5.33 \pm  2.11$ & $3.53 \pm 1.93$ & 0.41 & 14.00 \\
\hline
\multirow{2}{*}{$\bar{M}_{\star \textsc{mm}}$}   & OF & $0.30 \pm 0.20$ & $0.20 \pm  0.08$ & $0.32 \pm 0.21$ & 0.45 &  8.22 \\
                                                 & RF & $0.46 \pm 0.34$ & $0.27 \pm  0.17$ & $0.50 \pm 0.36$ & 0.54 &  2.82 \\
\hline
\multirow{2}{*}{$\bar{O}_{\textsc{sys-max}}$}    & OF & $2.05 \pm 1.00$ & $2.67 \pm  0.82$ & $1.88 \pm 0.96$ & 0.47 &  6.81 \\
                                                 & RF & $2.12 \pm 0.73$ & $2.44 \pm  0.68$ & $2.03 \pm 0.71$ & 0.21 & 86.28 \\
\hline
\multirow{2}{*}{$\bar{M}_{\textsc{sys-max}}$}    & OF & $0.45 \pm 0.24$ & $0.42 \pm  0.16$ & $0.47 \pm 0.25$ & 0.35 & 27.61 \\
                                                 & RF & $0.74 \pm 0.41$ & $0.60 \pm  0.31$ & $0.77 \pm 0.42$ & 0.37 & 21.49 \\
\hline
\end{tabular}
\end{table*}
%%%%%

%%%%%
\subsection{Initial mass function}\label{SEC:IMF}
%%%%%

The stellar IMF gives the probability that a newly formed star has a certain mass \citep{Chabrier2003}. The IMFs observed in different local star-forming regions appear to be very similar, implying that the star formation process is independent of environment \citep{Kroupa2001, Kroupa2002}. Numerical simulations reproduce this universal IMF well for a large variety of initial conditions \citep{Bate2005b, Bate2009a, Bate2009b, Bate2012}. 

\begin{figure}
\centering
\includegraphics[width=0.5\textwidth]{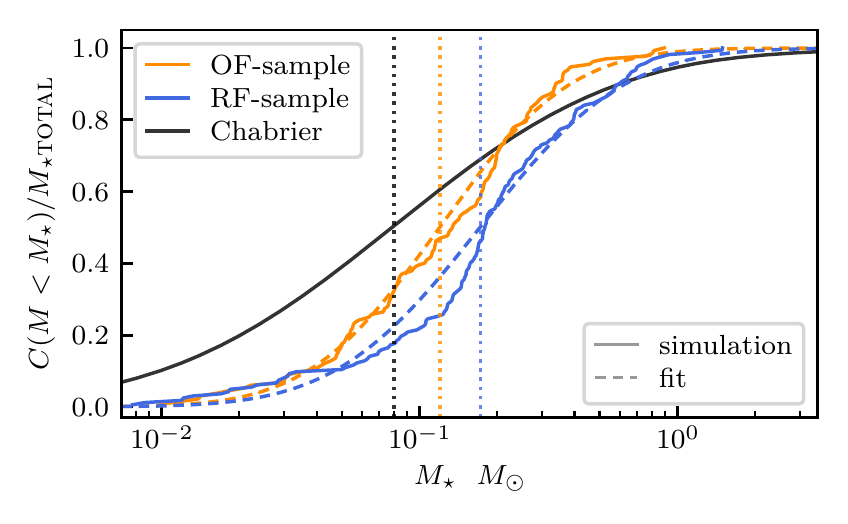}
\caption{\label{fig:IMF_Cum} Cumulative mass functions for the OF-sample (132 stars, solid orange line) and the RF-sample (163 stars, solid blue line) at $t_5=t_0+5\,\tff$. The dashed lines show log-normal fits to these distributions, and the dotted lines indicate the mean values, $\bar{M}\tsc{of} =0.13\,\Ms$ and $\bar{M}\tsc{rf}=0.18\,\Ms$. The black line shows the Chabrier IMF for comparison ($\bar{M}\tsc{chab} =0.08\,\Ms$).}
\end{figure}

We cannot attempt to reproduce the observed IMF here because we have only treated a single core mass ($1\,\Ms$). Observed cores are known to have a distribution of masses, given by the core mass function (CMF), and the CMF appears to be similar in shape to the IMF but shifted to higher masses \citep[e.g.][]{Andre10, Konyves15, Konyves20}. However we can evaluate the mean stellar mass function (MF) produced by a $1\Ms$ core, with and without outflow feedback. Fig.~\ref{fig:IMF_Cum} shows the Chabrier IMF (solid black line), the cumulative MF for the OF-sample (solid orange line), the cumulative MF for the RF-sample (solid blue line), a log-normal fit to the OF sample (dashed orange line) and a log-normal fit to the RF sample (dashed blue line). The fits are obtained using data-likelihood maximisation Markov-Chain Monte-Carlo sampling. The mean masses, $\bar{M}_{\textsc{chab}}=0.08\,\Ms$, $\,\bar{M}_{\textsc{of}}=0.13\,\Ms\,$ and $\,\bar{M}_{\textsc{rf}}=0.18\Ms\,$ are shown as vertical dotted lines. The corresponding standard deviations are $\sigma\tsc{chab}=0.69\pm 0.05$, $\sigma\tsc{of}=0.40\pm 0.06$ and $\sigma\tsc{rf}=0.44\pm 0.06$. Thus, outflow feedback reduces the mean stellar mass produced by a $1\Ms$ core by $\sim 28\%$ \citep[cf.][]{Krumholz12,Hansen12}. If we compare the cumulative MFs using the KS test, it returns statistics $s=0.24$ and $p<0.1\%$. We conclude that the OF- and the RF-samples are not drawn from the same underlying distribution. This conclusion is confirmed by an Anderson--Darling test \citep{Stephens74}.

If the fragmentation of a core into stars is a statistically self-similar process -- in the sense that the probability that a core of mass $M\tsc{core}$ spawns a star of mass $M_\star$ is the same as the probability that a core of mass $\beta M\tsc{core}$ spawns a star of mass $\beta M_\star$ -- the width of the observed IMF is
\begin{eqnarray}\label{EQN:sgma.1}
\sigma_{\textsc{chab}}&\simeq&\sqrt{\sigma_{\textsc{core}}^2 + \sigma_{\textsc{frag}}^2 } \, .
\end{eqnarray}
Here, $\sigma_{\textsc{core}}$ is the logarithmic standard deviation of the CMF, and $\sigma_{\textsc{frag}}$ is the logarithmic standard deviation of the stellar MF from a single core. Eqn. (\ref{EQN:sgma.1}) implicitly assumes that both the CMF, and the stellar MF from a single core, are approximately log-normal. Substituting $\sigma_{\textsc{frag}}=\sigma\tsc{of}$, we obtain
\begin{eqnarray}
\sigma_{\textsc{core}}&\simeq\sqrt{\sigma_{\textsc{chab}}^2-\sigma_{\textsc{of}}^2 }\;\,\simeq\;\,0.57\pm 0.07\,.\;\;\;
\end{eqnarray}
In other words -- if the assumption of statistically self-similar core fragmentation is correct -- the CMF makes a larger contribution to the standard deviation of the IMF than the process of core fragmentation. However, we should be mindful that the fragmentation of more massive cores might be very different from those we have simulated here.

%%%%%
%%%%%

%%%%%
\section{Multiplicity}\label{SEC:Multiplicity}
%%%%%
Most field stars with $M_\star\ga\Ms$, and a high fraction of those with lower mass, are in multiple systems \citep[e.g.][]{Raghavan10, Whitworth15}. The fraction of newly-formed stars in multiple systems is even higher, and the presumption is that some of these multiples are subsequently ionised by N-body interactions or tidal stresses to produce the distribution in the field. It follows that numerical simulations of star formation should (a) reproduce the multiplicity statistics observed, and (b) demonstrate how multiple systems actually form.

We use three multiplicity descriptors \citep[e.g.][]{Reipurth1993}. The multiplicity frequency,
\begin{eqnarray}\label{eq.MF}
m\!f = \frac{B_2+T_3+Q_4+Q_5+\,...}{S_1+B_2+T_3+Q_4+Q_5+\,...}\,,
\end{eqnarray}
gives the number of systems with more than one member (i.e. order higher than one). The higher-order frequency,
\begin{eqnarray}\label{eq.HF}
h\!f = \frac{T_3+Q_4+Q_5+\,...}{S_1+B_2+T_3+Q_4+Q_5+\,...} \, .
\end{eqnarray}
gives the number of systems with more than two members (i.e. order higher than two). The pairing factor, 
\begin{eqnarray}\label{eq.PF}
p\!f = \frac{B_2+2T_3+3Q_4+4Q_5+\,...}{S_1+B_2+T_3+Q_4+Q_5+\,...}\,, 
\end{eqnarray}
gives the average number of companions to a randomly picked primary star.

%%%%%
\subsection{VANDAM survey}\label{SEC:Vandam}
%%%%%
We compare the multiplicity statistics from our simulations with those from the {\sc vandam} survey \citep{Tobin16}, which used the VLA to measure the multiplicity statistics of 64 Class 0/I multiple protostars with separations between $15\,{\rm AU}$ and $10,000\,{\rm AU}$, in the Perseus molecular cloud. A proper comparison would require the generation of synthetic observations, taking account of sensitivity, beam size, UV-coverage, confusion and projection; for example, some of the close binary systems in our simulations have very small separations and might not be detectable as binaries. However, generating synthetic observations is outside the scope of this paper, and therefore we simply make direct comparisons between our simulations and the observations.

\begin{figure}
    \centering
    \includegraphics[width=0.5\textwidth]{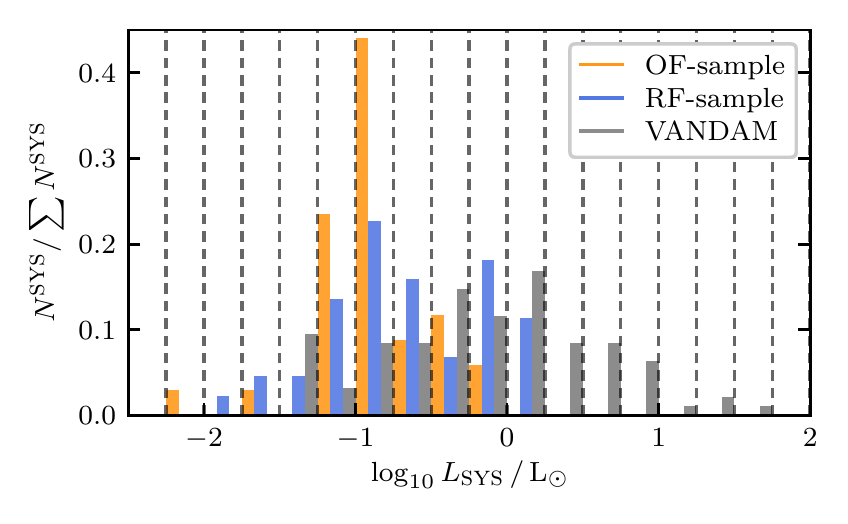}  \caption{\label{fig:LumDist}
    Stellar bolometric luminosities of all multiple systems in the OF- (orange) and RF-sample (blue) at $t_5$ compared to the observed bolometric luminosities of multiples in the {\sc vandam} survey (grey) \citep{Tobin16}. The luminosities of the {\sc vandam} survey multiples are higher, suggesting that somewhat more massive stars are present.}
\end{figure}

The protostars observed within the {\sc vandam} survey are slightly more massive than the stars in our OF-sample. \citet{Tobin16} do not provide masses for individual observed stars or multiple systems. However, using the protostellar luminosity function of \cite{McKee10,McKee11} they compute a protostellar mass function and expect their stars to be progenitors of K- and M-dwarfs (0.08 - 0.8 M$_{\odot}$) with a mean protostellar mass of $\sim 0.2 \, \Ms$ of which $ \sim 14 \, \%$ have masses between 0.7 $\Ms$ and 2.5 $\Ms$. In our simulations the mean protostellar mass at $t_5$ is 0.17 $\pm$ 0.15 M$_{\odot}$ (OF-sample) and 0.26 $\pm$ 0.25 M$_{\odot}$ (RF-sample), respectively. Only $3 \, \%$ (OF-sample) and $5 \, \%$ (RF-sample) of the stars are more massive than 0.7 $\Ms$. Fig. \ref{fig:LumDist} shows the stellar bolometric luminosity distribution of the multiple systems in our simulations at $t_5$. The luminosities are computed using the stellar evolution model by \cite{Offner09} (Section \ref{SEC:StellarEvolution} and \ref{SEC:RadFeedback}). Since, at this point, most of the gas is either bound in stars or entrained by the outflows, we expect the stellar bolometric luminosity to be comparable with those observed by \citet{Tobin16}. Comparing the luminosities of multiple systems in our simulations to the {\sc vandam} survey (Fig. \ref{fig:LumDist}) indicates that we are not probing exactly the same mass range. Despite this difference, the {\sc vandam} survey is still the best survey of protostellar multiple systems and the only one we can compare our simulation with.

%%%%
%%%%%

%%%%%
\subsection{Multiplicity statistics}\label{SEC:MultStatistics}
%%%%%

\begin{figure}
    \centering
    \includegraphics[width=0.5\textwidth]{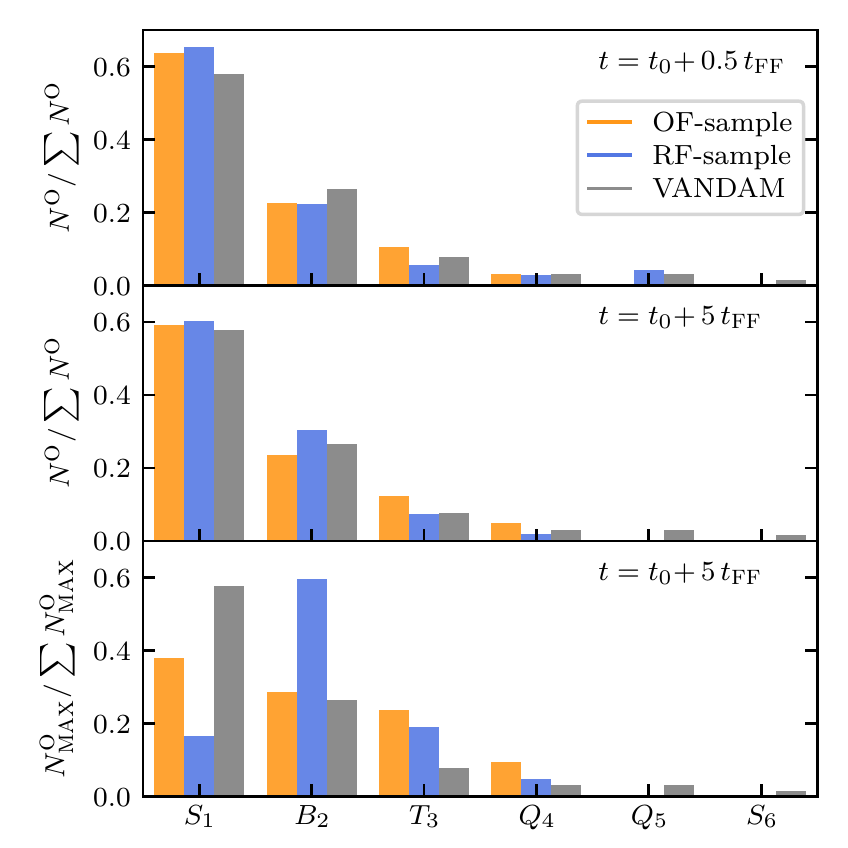}
    \caption{\label{fig:Multiplicity-Hist} The fractions of systems that are single, binary, triple, etc., for the OF-sample (orange) and the RF-sample (blue), at $t_{0.5}$ (top panel) and $t_5$ (middle and bottom panels), compared with the {\sc vandam} survey (grey). The top and middle panels include all systems, whereas the bottom panel includes only the highest-order system from each simulation. The OF-sample distribution changes little between the top and middle panels. In contrast, many of the quadruple and quintuple systems that form early in the RF-sample (top panel) quickly decay into binaries (middle panel); this is even clearer in the bottom panel where binaries dominate the distribution of highest-order systems for the RF sample.}
\end{figure}

Fig.~\ref{fig:Multiplicity-Hist} shows the fractions of systems that are single, binary, triple, etc., for the OF-sample (orange) and the RF-sample (blue), at $t_{0.5}$ (top panel) and $t_5$ (middle and bottom panels), compared with the {\sc vandam} survey (grey). The top and middle panels show the fractions of all systems, $N^O/\sum^O\!\!\left\{\!N^O\!\right\}$, with $N^O$ the number of systems of order $O\;(=S_1,\;B_2,\;T_3,\;Q_4,\;Q_5,\;{\rm etc.})$. The bottom panel instead shows the distribution of $N_{\textsc{max}}^O/\sum^O\!\!\left\{\!N_{\textsc{max}}^O\!\right\}$, where $N_{\textsc{max}}^O$ only takes account of the highest-order system from each simulation (Section \ref{SEC:StarFormation}).

Already by $t_{0.5}$ (Fig.~\ref{fig:Multiplicity-Hist}, top panel) the RF-sample includes many HOMs, including quintuples. However, these systems are very unstable, and by  $t_5$ (Fig.~\ref{fig:Multiplicity-Hist}, middle panel) they have decayed to binaries. When taking into account only the highest-order systems, binaries dominate the distribution (Fig.~\ref{fig:Multiplicity-Hist}, bottom panel).

In contrast, the multiplicity distribution of the OF-sample remains rather constant between $t_{0.5}$ and $t_5$. It is a monotonically decreasing function of the order, and matches the {\sc vandam} survey well; the highest-order systems are not dominated by binaries.

The fraction of singles that are the highest-order system is low (Fig.~\ref{fig:Multiplicity-Hist}, bottom panel) because many singles are low-mass stars that are ejected during the dynamical interactions that reduce HOMs to binary systems. This is particularly true for the RF-sample, where 89\% of singles are ejecta; for the OF-sample only 67\% of singles are ejecta.

%%%%
%%%%%

%%%%%
\subsection{Time evolution of the stellar multiplicity}\label{SEC:MultTimeEv}
%%%%%

\begin{figure}
    \centering
    \includegraphics[width=0.5\textwidth]{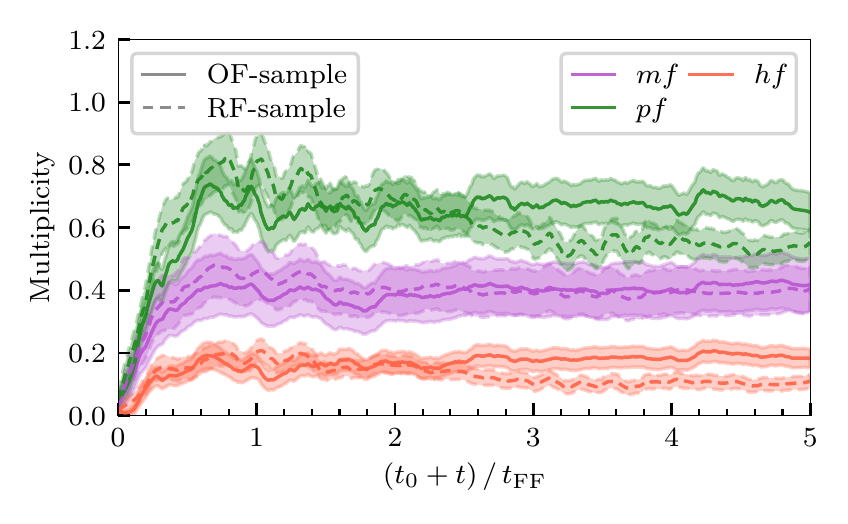}
    \caption{\label{fig:SF-Multiplicity} Time evolution of the multiplicity frequency, $m\!f$ (purple), pairing factor, $p\!f$ (green), and higher-order frequency, $h\!f$ (red), for the OF-sample (solid lines) and RF-sample (dashed lines). The areas, surrounding the lines in corresponding colours, indicate the time-dependent, propagated upper-limit on the Poisson uncertainty. The x-axis shows the time after the formation of the first sink, in units of the free fall time. While the multiplicity frequency is comparable for the two samples, the pairing factor and higher-order frequency are significantly higher for the OF-sample after $2.5\tff$.}
\end{figure}

Fig \ref{fig:SF-Multiplicity} shows the time evolution of the multiplicity descriptors defined in Eqns. (\ref{eq.MF}) through (\ref{eq.PF}), for all the simulations in the OF- and RF-samples. Here, time is measured from when the first sink forms ($t_0$), in units of the freefall time of the birth core ($\tff$). Once star formation starts, the multiplicity rises rapidly up to $\sim 0.6\tff$. Thereafter the multiplicity frequency is approximately constant and comparable for both samples, $m\!f_{_{\rm \!OF}}\sim m\!f_{_{\rm \!RF}}\sim 0.40$. For the OF-sample, most of the multiple systems are quite stable, and therefore the higher-order frequency and pairing factor are also approximately constant, at $h\!f_{_{\rm \!OF}}\sim 0.15$ and $p\!f_{_{\rm \!OF}}\sim 0.65$ respectively. However, for the RF-sample, the HOMs immediately start to eject lower-mass members and decay to binaries; this has no effect on the multiplicity frequency, $m\!f_{_{\rm \!RF}}$, but the higher-order frequency and pairing factor both decrease steadily, and by $t\sim 5\tff$, they are $h\!f_{_{\rm \!RF}}\sim 0.10$ and $p\!f_{_{\rm \!RF}} \sim 0.50$. 

The multiplicities of the OF- and RF-samples are very similar in the early phase of star formation. The RF-sample forms somewhat more multiples and forms them somewhat faster, but these differences are small, and most of the HOMs formed in the RF-sample quickly reduce to binaries.  HOMs are more stable against disruption when outflows are present.

Table \ref{Table:Multiplicity} compares $m\!f$, $p\!f$ and $h\!f$ for both samples at $t_5$ with the {\sc vandam} survey. All the statistics for the OF-sample agrees with the {\sc vandam} survey within the uncertainties; $m\!f_{_{\rm OF}}$ and $h\!f_{_{\rm OF}}$ agree very well, but $p\!f_{_{\rm OF}}$ only just agrees. For the RF-sample, only $m\!f_{_{\rm RF}}$ agrees with {\sc vandam}, both $p\!f_{_{\rm RF}}$ and $h\!f_{_{\rm RF}}$ are much too low. This is largely due to the decay of HOMs in the RF-sample.

Note that a direct comparison with the {\sc vandam} survey is biased since they observe slightly more massive stars than we produce in our simulations (Section \ref{SEC:Vandam}). Observations show that the multiplicity fraction is strongly dependent on the primary mass  \citep[see, e.g. Fig. 1 in][]{Whitworth15}. However, this relation is valid for main-sequence stars and it is not clear whether it holds for the pre-main-sequence regime. Taking this relation into account, it seems that our simulations produce a too high multiplicity, given the low protostellar masses. However, our simulations do not include magnetic fields. With magnetic fields, we would expect less fragmentation and hence a somewhat lower multiplicity (see the discussion in Section \ref{SEC:MagneticDiscs}). The missing magnetic fields could possibly explain why we are matching the {\sc vandam} survey so well, even though we are probing a lower mass regime.

%%%%%
%%%%%

% the abscissa should be labelled '$(t-t_0)/\tff$'

%%%%%
\begin{table}
\caption{Multiplicity statistics for the OF- and RF-samples at $t_5$, compared with the VANDAM survey \citep{Tobin16}. The first column gives the sample, followed by the multiplicity frequency (Eq. \ref{eq.MF}), the pairing factor (Eq. \ref{eq.PF}) and the high-order frequency (Eq. \ref{eq.HF}).} 
\begin{center}\begin{tabular}{lccc}
\hline 
Sample & $m\!f$ & $p\!f$  & $h\!f$   \\
\hline 
RF-sample    & $0.40 \pm 0.07$ & 0.51 $\pm 0.05$ & 0.09 $\pm 0.02$ \\
OF-sample    & $0.40 \pm 0.09$ & 0.63 $\pm 0.06$ & 0.17 $\pm 0.03$ \\
VANDAM       & $0.40 \pm 0.06$ & $0.71 \pm 0.06$ & 0.16 $\pm 0.04$ \\
\hline
\end{tabular}\end{center} 
\label{Table:Multiplicity}
\end{table}
%%%%%

%%%%%
\subsection{Stability of triple systems}\label{SEC:Stability}
%%%%%

Figure \ref{fig:SF-Multiplicity} shows that the fraction of HOMs in the RF-sample decreases after $t_{1}$, while the fraction is almost constant for the OF-sample. To confirm objectively that this is because the hierarchical triple systems in the OF-sample are more stable, we compute the two different criteria for the stability of hierarchical triple systems which \cite{Zhuchkov10} has shown to be most reliable. 

An hierarchical triple system is one in which a pair of stars (labelled individually `1' and `2', and together `1+2') are on a tight orbit around one another, and this pair and a third star (labelled `3') are then on a much wider orbit around one another. The masses of the stars are $m_1$, $m_2$, and $m_3$, and the total mass of the tight pair is $m\tsc{in}\!=\!m_1\!+\!m_2$. The semi-major axis and eccentricity of the tight orbit (involving stars 1 and 2) are $a\tsc{in}$ and $e\tsc{in}$. The semi-major axis and eccentricity of the wide orbit (involving star 3 and the pair 1+2) are $a\tsc{out}$ and $e\tsc{out}$. The system is hierarchical in the sense that $a\tsc{out}\gg a\tsc{in}$. With these definitions, the criterion for stability developed by \cite{Aarseth03} is
\begin{eqnarray}\label{EQN:Aarseth}
f_{\textsc{a}}\!\!\!&\!\!=\!\!&\!\!0.36\,\frac{a_{\textsc{out}} (1\!-\!e_{\textsc{out}})}{a_{\textsc{in}}}\!\left[\!\left(\!1\!+\!\frac{m_{3}}{m\tsc{in}}\!\right)\!\frac{1 + e_{\textsc{out}}}{\sqrt{1 - e_{\textsc{out}}}}\right]^{-2/5}>1;\hspace{0.8cm}
\end{eqnarray}
and the criterion developed by \cite{Valtonen07} is
\begin{eqnarray}\nonumber
f_{\textsc{v}}\!\!&\!\!=\!\!&\!\!3\,\frac{a_{\textsc{out}} (1\!-\!e_{\textsc{out}})^{7/6}}{a_{\textsc{in}}}\!\left[\!\left(\!1\!+\!\frac{m_{3}}{m\tsc{in}}\!\right)\!\left(\!\frac{7}{4}\!+\!\frac{\cos(i)}{2}\!-\!\cos^2(i)\!\right)\right]^{-1/3}\\\label{EQN:Valtonen}
\!\!&\!\!>\!\!&\!\!1,
\end{eqnarray}
where  $i$ is the angle between the angular momentum vectors of the tight and wide orbits. 

\begin{figure}
    \centering
    \includegraphics[width=0.499\textwidth]{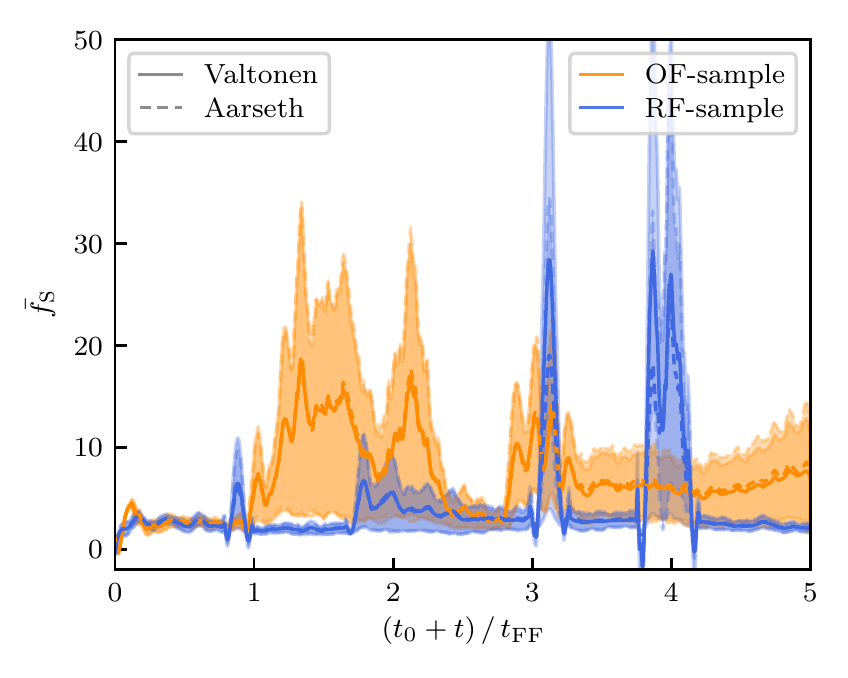}
    \caption{Time evolution of the Aarseth and Valtonen stability parameters (Eqns. \ref{EQN:Aarseth} and \ref{EQN:Valtonen}), averaged over all the triple systems in the OF-sample (orange), and over all the triple systems in the RF-sample (blue). The dashed lines show the results obtained for the Aarseth parameter ($f\tsc{a}$), and the solid lines show the results obtained for the Valtonen parameter ($f\tsc{v}$). At most times the dashed lines cannot be discerned because they sit on top of the solid lines. The coloured shading represents the standard deviation about the mean. Triple systems in the OF-sample are markely more stable than triple systems in the RF-sample.}
   \label{fig:StabilityTimeEv}
\end{figure}

Fig \ref{fig:StabilityTimeEv} shows the time evolution of $\bar{f}\tsc{a.of}$ (orange dashed line) and $\bar{f}\tsc{v.of}$ (orange full line), i.e. $f\tsc{a}$ and $f\tsc{v}$ averaged over all the triple systems in the O-sample; and the time evolution of $\bar{f}\tsc{a.rf}$ (blue dashed line) and $\bar{f}\tsc{v.rf}$ (blue full line), i.e. $f\tsc{a}$ and $f\tsc{v}$ averaged over all the triple systems in the R-sample. $\bar{f}\tsc{a.of}$  is almost indistinguishable from $\bar{f}\tsc{v.of}$, and likewise $\bar{f}\tsc{a.rf}$ from $\bar{f}\tsc{v.rf}$, indicating that they are mutually consistent. After $t\tsc{ff}$, $\bar{f}\tsc{a.of}$, and $\bar{f}\tsc{v.of}$ are almost always well above $\bar{f}\tsc{a.rf}$ and $\bar{f}\tsc{v.rf}$, on average by a factor $2.53 \pm 0.05$. $\bar{f}\tsc{a.of}$ and $\bar{f}\tsc{v.of}$ are also well above unity most of the time, and end up at $\sim 7$, so most of the triples are stable. In contrast, $\bar{f}\tsc{a.rf}$ and $\bar{f}\tsc{v.rf}$ are almost always $\sim 2$, so outliers with lower than average values tend to be unstable and decay.

%%%%%
%%%%%

%%%%%
\subsection{Twin-binaries}\label{SEC:Twin}
%%%%%

\begin{figure}
    \centering
    \includegraphics[width=0.5\textwidth]{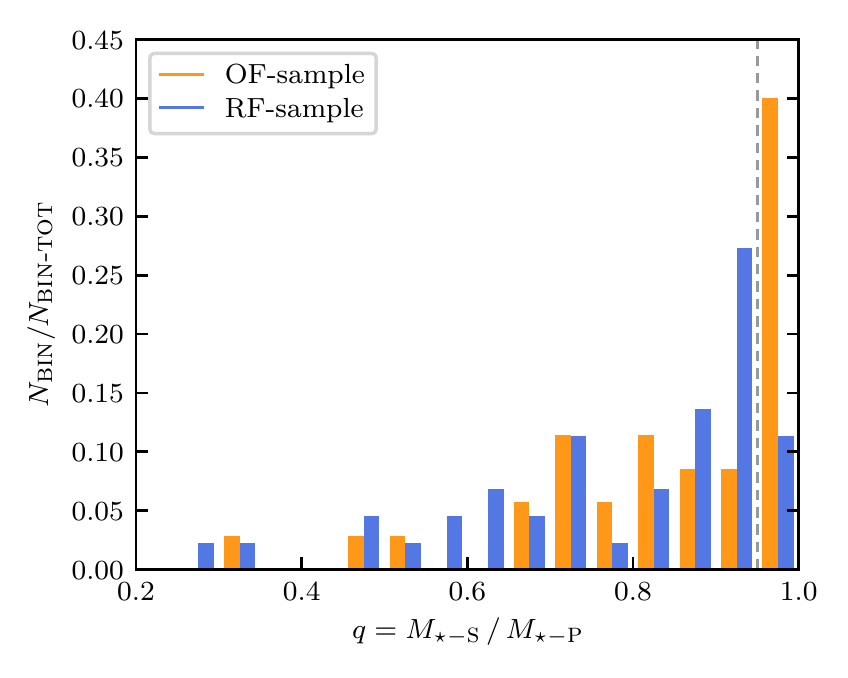}
    \caption{\label{fig:TwinBinaries} The normalised distribution of the mass ratio between the secondary and primary components, $q= M_{\star-\textsc{s}}/M_{\star-\textsc{p}}$, for all binary systems in the OF-sample (orange) and all binary systems in the RF-sample (blue), at $t_5\!\equiv\!t_0\!+\!5t\tsc{ff}$. At this stage there are 35 binaries in the OF-sample, and 44 in the RF-sample. Both distributions are peaked at high $q$, with a tail towards low $q$. Outflow feedback appears to enhance the formation of twin binaries with $q \ge 0.95$ (bin to the right of the dashed grey line). }
\end{figure}

Observations of low-mass binary systems reveal a high fraction of systems in which the ratio of the secondary mass to the primary mass, $q = M_{\star-\textsc{s}}/M_{\star-\textsc{p}}$, is close to unity \citep[see][or the review by \cite{Duchene13}]{Lucy06,Simon09,Fernandez17,Kounkel19,El-Badry19}; these systems are referred to as `twin binaries'. Fig \ref{fig:TwinBinaries} shows the distribution of $q$ for the 35 binary systems in the OF-sample, and the 44 binaries in the RF-sample. Both samples are peaked at high $q$ with a tail towards low $q$, similar to the observations reported by \cite{Fernandez17} and \cite{Kounkel19}. The convention \citep[e.g.][]{Kounkel19, El-Badry19} is to class binary systems with $q \ge q\tsc{crit}=0.95$ as twins. Using this definition, $f_\textsc{twin-of}=43\%$ for the OF-sample, compared with $f_\textsc{twin-rf}=9\%$ for the RF-sample. Since, with the small number of binaries in our sample, the ratio $f_\textsc{twin-of}/f_\textsc{twin-rf}$ is particularly high for $q\tsc{crit}=0.95$, we have varied $q\tsc{crit}$ between $0.90$ and $0.99$, but for all $q$-values in this range, there are always more twins in the OF-sample, $f_\textsc{twin-of}/f_\textsc{twin-rf}\geq 2.1\pm 0.6$.

The excess of almost equal mass binaries is thought to be the result of competitive accretion \citep{Tokovinin00}. For a low-$q$ binary, the secondary is on a larger orbit and therefore accretes matter with high specific angular momentum faster than the primary, thereby driving the mass ratio towards unity \citep{Whitworth95,Young15, Matsumoto19}. In the present context this happens because the binary is accreting from a circumbinary disc \citep[see Fig. \ref{fig:ColDens} and][]{El-Badry19}.

%%%%%
%%%%%

\begin{figure*}
    \centering
    \includegraphics[width=0.99\textwidth]{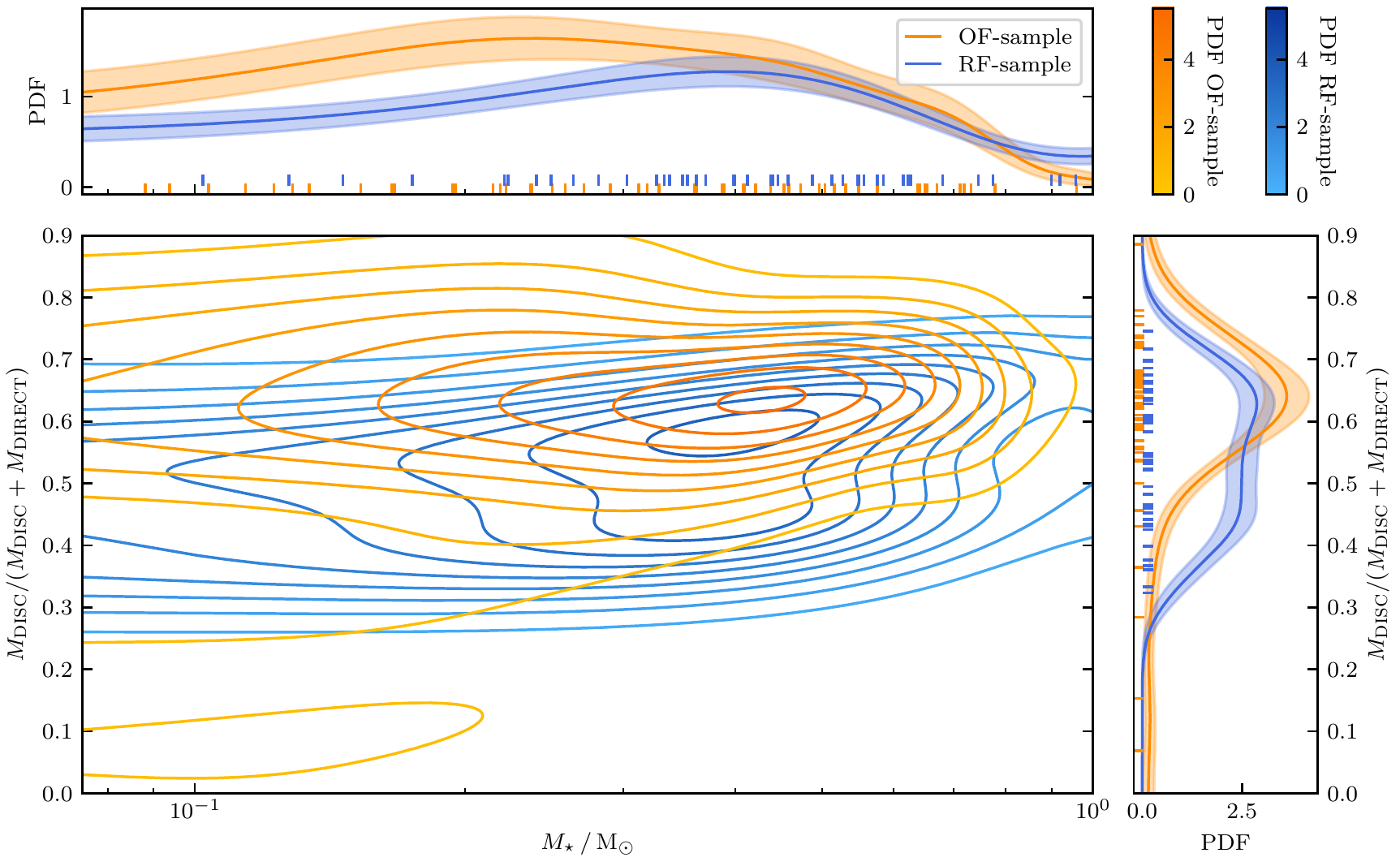}
    \caption{\label{fig:2dKDE} The two-dimensional probability density function (PDF) for the `final' stellar mass, $M_\star$, against the fraction of that mass that has been acquired by disc accretion (as opposed to by direct infall), $M\tsc{disc}/(M\tsc{disc}+M\tsc{infall})$; here `final' means at $t_{5}\!\equiv\!t_0\!+\!5t\tsc{ff}$. Contours at 10\%, 20\%, . . . . . 90\% of the maximum probability are in shades of orange for the OF-sample, and in shades of blue for the RF-sample. The top panel shows the one-dimensional PDF for $M_\star$. The right panel shows the one-dimensional PDF for $M\tsc{disc}/(M\tsc{disc}+M\tsc{infall})$. On the top and right panels, the shaded regions indicate uncertainties estimated using bootstrapping, and the small tick marks represent individual systems. Evidently simulations with outflow feedback lead to stars accreting more gas by disc accretion.}
\end{figure*}

Figure \ref{fig:2dKDE} shows the two-dimensional probability density function (PDF) for the stellar mass $M_{\star}$ at $t_5$ and the fraction of this mass acquired by disc accretion as opposed to direct infall, $f_{\textsc{disc}} = M_{\textsc{disc}} / \left(M_{\textsc{disc}} + M_{\textsc{direct}} \right)$. The corresponding one-dimensional PDFs are shown in the top and right panels. 

To compute $f_{\textsc{disc}}$ we evaluate the mass flow through a sphere around each star, with the radius of the sphere equal to half the radius of the corresponding accretion disc. SPH particles are assigned to an accretion disc if (a) their density is $>10^{-14} \gcm$, (b) their rotational velocity component is greater than their radial velocity component with respect to the corresponding sink, and (c) they are gravitationally bound to the star. The mean disc radius for the OF-sample at $t_2$ is $\bar{r}_{\textsc{disc}} = 103 \pm 55$ AU, which is in good agreement with recent observations of Class 0/I stars \citep{Maury19}. Only stars in multiple systems that have both an accretion disc and a velocity lower than the escape speed contribute to the evaluation of $f\tsc{disc}$. The mass flow contributes to $M_{\textsc{disc}}$ if $\frac{\pi - \theta_{\tsc{open}}}{2} < \theta < \frac{\pi + \theta_{\tsc{open}}}{2} $, where $\theta$ is the angle between the position vector (relative to the star) and the angular momentum axis of the disc. We set $\theta_{\tsc{open}} = \mathrm{1/3} \, \pi$; varying $\theta_{\tsc{open}}$ between 1/2 $\pi$ and 1/6 $\pi$ shows no qualitative difference.

The peak of the OF-sample PDF is shifted to higher $f_{\textsc{disc}}$ compared with the peak of the RF-sample. A one-sided Mann-Whitney-U test confirms that this difference is significant with $p\ll 1\%$. Outflow cavities stop stars from accreting so rapidly via direct infall, thereby enhancing the contribution from disc accretion and hence increasing the fraction of twin binaries in the OF-sample.

%%%%%
%%%%%

%%%%%
\subsection{Magnetic fields and disc fragmentation}\label{SEC:MagneticDiscs}
%%%%%

Observations show that dense cores are threaded by magnetic fields \citep{Troland08, Kandori18}. Recent numerical simulations suggest that magnetic fields have a profound impact on the star formation process \citep[see, e.g. the review of][]{Wurster18}. In the ideal magnetohydrodynamic (MHD) case, magnetic braking almost entirely prohibits the formation of accretion discs \citep{Commercon12, Bate14}. However, the introduction of non-ideal MHD effects mitigates the efficiency of magnetic braking to some extent \citep{Hennebelle16, Wurster16, Zhao18}. Moreover, turbulence can cause a misalignment of the angular momentum and the magnetic field vector, which may significantly reduce the magnetic braking efficiency and enable the formation of massive discs \citep{Seifried13, Seifried15, Wurster16, Gray18, Wurster18, Wurster19b}. Including the Hall-effect in non-ideal MHD simulations, \citet{Wurster18} show that for the case of anti-aligned magnetic field and angular momentum vectors, they obtain results which are most similar to a pure hydrodynamical calculation. The magnetic Toomre-Q parameter implies that these magnetized discs are generally more stable against fragmentation than pure hydrodynamical discs \citep{Toomre64, Kim01, Wurster19a}.  

Our code, {\sc Gandalf}, does not currently include magnetic fields. With magnetic fields, we would expect to have slightly smaller discs that fragment less readily \citep{Hennebelle19, Wurster19a}. Therefore, the number of simulations forming a single star would increase resulting in fewer ejected stars and somewhat lower multiplicities.

Compared to the RF-sample, our OF-sample contains significantly more cores which form only a single star (Section \ref{SEC:StarFormation}). With episodic accretion feedback, the accretion discs are frequently severely disturbed and the amount of fragmentation is damped to a realistic level \citep{Stamatellos12a, Lomax14, Lomax15, Mercer17}. We therefore speculate that magnetic fields would have a limited additional effect on disc fragmentation if episodic accretion feedback were taken into account.

%%%%%
\section{Self-regulation of outflow feedback}\label{SEC:SelfReg}
%%%%%

%%%%%
\subsection{Entrainment factor}\label{SEC:Entrainment}
%%%%%            

Molecular outflows from protostars consist mainly of secondary entrained material \citep{Tabone17, Zhang19}, i.e. core gas that is swept up by the primary ejected gas. For low-mass star formation the entrained gas mass is estimated to range from $0.1\,\Ms$ to $1.0\,\Ms$ \citep{Arce07}. The entrainment factor $\epsilon_{\textsc{of}}$ is defined as the ratio of total outflowing mass, $M_{\textsc{out}}$, to primary ejected mass, $M_{\textsc{eject}}$. An SPH particle contributes to $M_{\textsc{out}}$ if its radial velocity is higher than the local escape velocity and at least $0.1\,\kms$. Using numerical magneto-hydrodynamic simulations, \cite{Offner17} conclude that $\epsilon_{\textsc{of}} \sim 4$. 

\begin{figure*}
    \centering
    \includegraphics[width=0.999\textwidth]{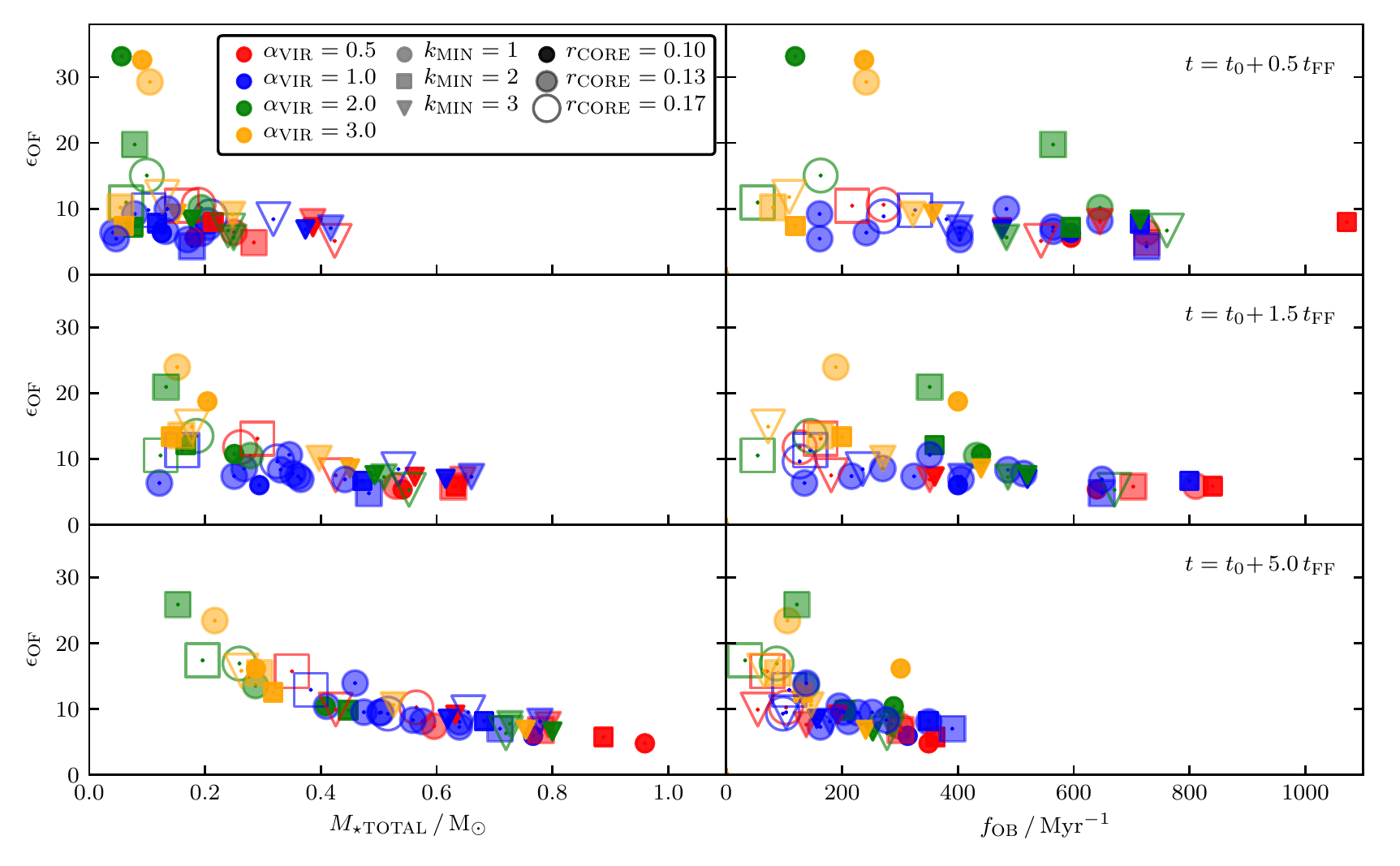}
    \caption{\label{fig:Entrainment} Entrainment factors, $\epsilon_{\textsc{of}}=M_{\textsc{out}}/M_{\textsc{eject}}$, for all the simulations in the OF-sample, plotted against total stellar mass, $M_{\star \textsc{total}}$ (left column), and against outburst frequency, $f_{\textsc{ob}}$ (right column), at $t_{0.5}$ (top row), $t_{1.5}$ (middle row), and $t_5$ (bottom row), where $t_\tau\!=\!t_0\!+\!\tau t\tsc{ff}$. At $t_5$, the entrainment factors range from $\epsilon_{\textsc{of}}\simeq 5$ to $\epsilon_{\textsc{of}}\simeq 26$. With increasing time, $\epsilon_{\textsc{of}}$ becomes strongly anti-correlated with $M_{\star \textsc{total}}$, and weakly anti-correlated with $f_{\textsc{of}}$.}
\end{figure*}

Fig.~\ref{fig:Entrainment} shows the entrainment factors at $t_{0.5}$ (top row), $t_{1.5}$ (middle row), and $t_5$ (bottom row), plotted against the total stellar mass, $M_{\star \textsc{total}}$ (left column) and against outburst frequency, $f_{\textsc{ob}}$ (right column). Most simulations have $\epsilon_{\textsc{of}} \sim 7$, but some have much higher values, up to $\epsilon_{\textsc{of}} \sim 33$, particularly at low $M_{\star \textsc{total}}$ and/or low $f_{\textsc{ob}}$. By $t_5$ there is a well-defined anti-correlation between $\epsilon_{\textsc{of}}$ and $M_{\star \textsc{total}}$, with $5\la\,\epsilon_{\textsc{of}}\la\,26$ and Spearman Rank Correlation coefficient $r_{\textsc{s}}=-0.94$ and $p\ll 0.01\%$. 

There are several possible reasons for this anti-correlation. The higher the total mass in stars, the less mass there is left in the core envelope, and therefore the less mass there is left to entrain. Moreover, as time advances the outflows are increasingly likely to be launched into cavities blown by previous outflows, in which case there is even less material for them to entrain; this is especially true when a core has formed a multiple system with aligned outflows. 

The right column of Fig.~\ref{fig:Entrainment}, shows the entrainment factor, $\epsilon_{\textsc{of}}$, against the outburst frequency, $f_{\textsc{ob}}$, averaged between the time when the first star forms, $t_0$, and -- respectively -- $t_{0.5}$, $t_{1.5}$ and $t_5$. There is an anti-correlation between $\epsilon_{\textsc{of}}$ and $f_{\textsc{ob}}$ which gets stronger with time. By $t_5$, the Spearman rank correlation coefficient is $r=-\,0.70$ with $p\ll 0.01\%$. Simulations with lower $f_{\textsc{ob}}$ (i.e. episodic accretion and outflow concentrated in a few massive outbursts) have higher $\epsilon_{\textsc{of}}$. The asymptotic limit of high $f_{\textsc{ob}}$ is continuous outflow, and this might explain the lower $\epsilon_{\textsc{of}}\sim 4$ found by \cite{Offner17}, since the protostars in their simulations generate continuous outflows.

%%%%%
%%%%%

%%%%%
\subsection{Outflowing gas mass}\label{Section:Mout}
%%%%%

Fig.~\ref{fig:SFE} (right column) shows the outflowing gas mass, $M_{\textsc{out}}$, for each simulation in the OF-sample, plotted against its total stellar mass, $M_{\star \textsc{total}}$, at $t_{0.5}$, $t_{1.5}$ and $t_5$. At $t_{0.5}$, $M_{\textsc{out}}$ and $M_{\star \textsc{total}}$ are tightly correlated, $M_{\textsc{out}}\simeq 0.01\Ms +0.58M_{\star \textsc{total}}$, with Spearman Rank Correlation coefficient $r=0.76$ and $p\ll 0.01\%$. Since $10\%$ of the matter entering a sink is ejected, this corresponds to an average entrainment factor of $\epsilon_{\textsc{of}} \sim 0.58/0.1\sim 6.4$, in good agreement with Fig.~\ref{fig:Entrainment}. 

At later times, the correlation remains strong but flattens, because cores with low $M_{\star \textsc{total}}$ have more gas left, and therefore higher entrainment factors and more massive outflows (see Fig.~\ref{fig:Entrainment}). By $t_5$, the correlation has become $M_{\textsc{out}}\simeq 0.29\Ms+0.29M_{\star \textsc{total}}$ with a non-zero intercept.

%%%%%
\subsection{Relative star formation efficiency}
%%%%%

\begin{figure*}
    \centering
    \includegraphics[width=0.999\textwidth]{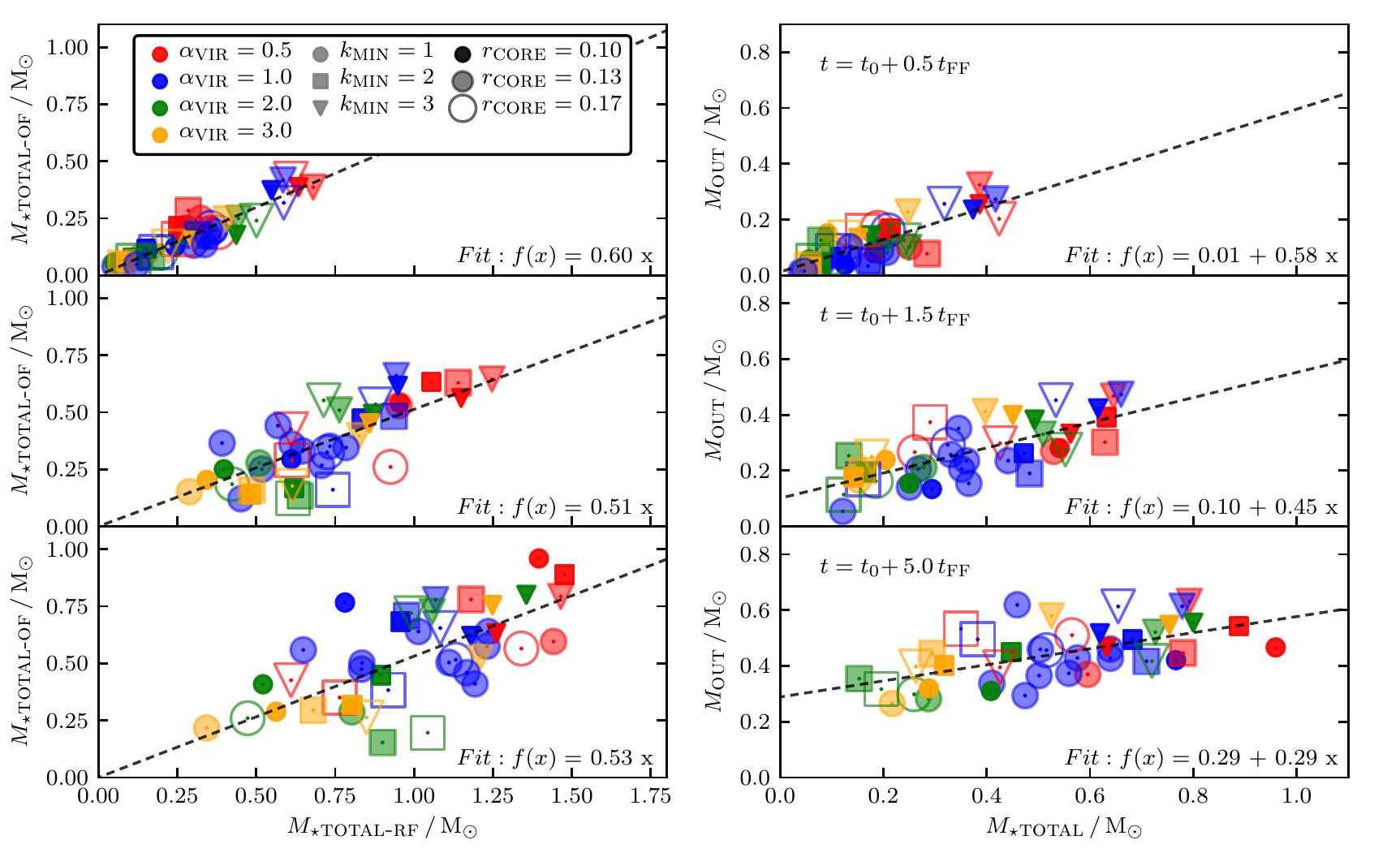}
    \caption{\label{fig:SFE}
    {\it Left:} The total stellar mass for each simulation in the OF-sample against the total stellar mass for the corresponding simulation in the RF-sample, at $t_{0.5}$ (top), $t_{1.5}$ (middle) and $t_5$ (bottom), where $t_\tau\!=\!t_0\!+\!\tau t\tsc{ff}$. The dashed lines are linear fits to the data, with $0.5\la\,M_{\star \textsc{total-of}}/M_{\star \textsc{total-rf}}\la\,0.6$. At $t_{0.5}$ the fit is very tight, but the scatter around the fit increases at later times, due to changes in the entrainment factor and hence changes in the effectiveness of outflow feedback (see Fig.~\ref{fig:Entrainment}). {\it Right:} The outflowing gas mass, $M_{\textsc{out}}$, for all the simulations in the OF-sample, against the total stellar mass, $M_{\star \textsc{total}}$, at $t_{0.5}$ (top), $t_{1.5}$ (middle) and $t_5$ (bottom). At $t_{0.5}$, there is a strong correlation, $M_{\textsc{out}}\simeq 0.01\Ms +0.58M_{\star \textsc{total}}$. At later times, the correlation remains strong, but flattens, becoming $M_{\textsc{out}}\simeq 0.29\Ms+0.29M_{\star \textsc{total}}$ at $t_5$; this is because by this stage cores with relatively low $M_{\star \textsc{total}}$ have more gas left, and therefore higher entrainment factors and -- relatively speaking -- more massive outflows (see Fig.~\ref{fig:Entrainment}).}
\end{figure*}

Fig.~\ref{fig:SFE} (left column) shows the total stellar mass of each simulation in the OF-sample, $M_{\star \textsc{total-of}}$, plotted against the total stellar mass in the corresponding simulation in the RF-sample, $M_{\star \textsc{total-rf}}$, at $t_{0.5}$, $t_{1.5}$ and $t_5$. At $t_{0.5}$, the masses are strongly correlated, $M_{\star \textsc{total-of}}\sim 0.60\,M_{\star \textsc{total-rf}}$, with Spearman Rank Correlation coefficient $r=0.90$ and $p\ll 0.01\%$. Thus, the early star formation rate (SFR) is $ \sim 40\%$ lower if outflows are present.

This strong correlation arises because (a) $f\tsc{eject}=0.1$, i.e. exactly 10\% of the matter entering a sink is ejected and the remaining 90\% is accreted, so $\left.dM_\star/dt\right|\tsc{of}=9\left.dM/dt\right|\tsc{eject}$; and (b) in the early stages the entrainment factor is approximately universal, $\epsilon\tsc{of}\simeq 7$, so $\left.dM/dt\right|\tsc{out}\sim 7\left.dM/dt\right|\tsc{eject}$. It follows that
\begin{eqnarray}
\frac{\left.dM_\star/dt\right|\tsc{of}}{\left.dM_\star/dt\right|\tsc{of}+\left.dM/dt\right|\tsc{out}}&\sim&\frac{9}{9+7}\;\,=\;\,56\%\,.\hspace{1.0cm}
\end{eqnarray}
If one assumes that, in the absence of outflow feedback, the outflowing matter would have ended up in the stars, the mass of stars in the OF-sample should be of order 56\% of the mass of stars in the RF-sample.

At later times the correlation persists, but with greater scatter. For example, at $t_5$, $M_{\star \textsc{total-of}}\simeq 0.53M_{\star \textsc{total-rf}}$ with Spearman Rank Correlation coefficient $r=0.65$ and $p< 0.01\%$.

%%%%%
%%%%%

%%%%%
\subsection{Possible effects of magnetic fields on the outflow structure}\label{SEC:MagneticOutflows}
%%%%%

Magnetic fields are obviously very important for protostellar outflows since magnetic fields are necessary to launch outflows in the first place \citep{Bally16}. This regime is covered by our episodic outflow sub-grid model, which is based on the episodic MRI instability of the inner disc. Computing the outflow launching self-consistently is not expected to alter the outflow properties significantly with respect to our sub-grid model \citep[see, e.g.][]{Federrath14}.

The influence of magnetic fields on the already launched or entrained gas is not well understood. Outflows are highly connected to the stellar accretion rate. Therefore, it is complicated to disentangle (i) the lower accretion rate due to magnetic fields \citep{Offner17} from (ii) the direct effects of magnetic fields on the outflowing gas. We argue that (i) episodic outflows are highly self-regulated \citep{Rohde18} and that therefore a slower collapse would not alter the outflow properties notably. \cite{Offner17} show that magnetic fields do not affect the entrainment factor (Section \ref{SEC:Entrainment}) and therefore (ii) the influence on already launched gas is limited. Moreover, the typical velocity of outflowing gas, up to $\sim 100 \, \kms$, is much higher than the characteristic Alfv\'en speed, $\upsilon\tsc{A} = B / \sqrt{\rho } = 1 \, \kms$ for typical values of $B = 10  \, \mu \mrm{G}$ and $\rho = 10^{-20} \, \gcm$ for the low-density gas in the outflow cavity. Therefore, we argue that including magnetic fields would not alter the properties of the outflows in this work significantly.

However, in simulations of star cluster formation with initial gas masses of 100 -- 1000 $\Ms$, it has been shown that the combination of outflows and magnetic fields is important. The momentum delivered by the outflows coupled to the magnetic fields maintains turbulence such that the parental molecular clouds stay close to virial equilibrium \citep[see the review by][]{Krumholz19}. Therefore, we might not be able to expand our simulations to larger scales without taking into account the potentially significantly higher impact of magnetic fields.

%%%%%
\section{Conclusions}\label{SEC:Conclusion}
%%%%%

Outflows are the dominant feedback mechanism in the early phase of low mass star formation. However, the consequences of outflow feedback for the evolution of a prestellar core are not well understood. Three questions are especially important. (1) Does outflow feedback affect the properties of the individual stars formed? (2)  Does outflow feedback affect the stellar multiplicity statistics? (3) How much does outflow feedback reduce the star formation efficiency?

To answer these questions, we have performed a large ensemble of smoothed particle hydrodynamics simulations of dense prestellar cores. All the cores have the dimensionless density profile of a Bonnor--Ebert sphere, but we vary the initial core radius ($\rc = 0.017\,\pc,\;0.013\,\pc,\;0.01\,\pc$), the initial virial ratio ($\avir = 0.5, \;1.0, \;2.0,\;3.0$), the wavenumber of the dominant mode in the initial turbulent velocity field ($\kmin = 1,\;2,\;3$). For each combination of $(\rc,\avir,\kmin)$ and eight additional runs with different turbulent seeds, we perform one simulation with outflow feedback, and one without. The resulting ensemble of 88 simulations reveals the following features.

\begin{itemize}
\item{The stellar statistics (total mass in stars, number of stars, mass of most massive star, total mass and order of the highest-order system) depend only weakly on the initial conditions of the birth core, i.e. $(\rc,\avir,\kmin)$.}
\item{The total mass in stars, the mass of the most massive star and the total mass of the highest-order system all tend to decrease markedly when outflow feedback is included. The total number of stars and the order of the highest-order system tend to decrease very slightly when outflow feedback is included -- except for the cores with low $\avir$ where these trends are reversed.}
\item{The distribution of stellar masses can be represented by a lognormal. Without outflow feedback the mean and standard deviation of $\log_{_{10}}(M_\star/\Ms)$ are $-0.74$ and $0.44$. When outflow feedback is included, they become $-0.89$ and $0.40$, i.e. the mean mass is reduced by $\sim 30\%$.}

\item{The simulations without outflow feedback produce a large number of higher-order multiples (HOMs), but many of them are unstable and quickly decay to binaries. As a result, these simulations deliver an higher-order frequency ($h\!f$) and a pairing factor ($p\!f$) that are inconsistent with the values observed in the {\sc vandam} survey by \citet{Tobin16}.}
\item{The simulations with outflow feedback produce slightly fewer HOMs, but most of them are stable. As a result, these simulations deliver an $h\!f$ and a $p\!f$ which, within the uncertainties, agree with the distribution observed in the {\sc vandam} survey by \citet{Tobin16}.}
\item{The inclusion of outflow feedback increases considerably the fraction of twin binaries with almost equal-mass components. This is because outflow feedback reduces the role of direct infall onto a growing protostar, so the components of a binary system have to acquire their mass by accretion from a circumbinary disc.}
\item{The mean entrainment factor (the ratio between outflowing mass at large radius and the mass ejected from the protostar and its disc) is $\bar{\epsilon}_{\textsc{of}}\sim 7$, significantly larger than the value of 4 obtained by \cite{Offner17} in simulations with {\it continuous} outflow. Above average $\epsilon_{\textsc{of}}$ values are confined to cores with low total stellar mass (especially those with high $\avir$ and low $\rc$), because there is then more gas left to entrain (and it is more easily unbound).}
\item{In the early stages, the outflowing mass, $M\tsc{out}$, is approximately proportional to the total stellar mass, $M\tsc{out}\sim 0.7M_{\star \textsc{total}}$. At later times this is still true in cores where $M_{\star \textsc{total}}$ is low, but in cores where $M_{\star \textsc{total}}$ is high, and there is less gas left to push out, $M\tsc{out}< 0.7M_{\star \textsc{total}}$.}
\item{Since we have assumed that the rate of mass ejection from a protostar is exactly proportional to the rate of mass accretion onto the protostar, and since there is not a huge variation in entrainment factors, the mass converted into stars when outflow feedback is included is an approximately constant fraction of the mass converted into stars when there is no outflow feedback; after 5 freefall times this fraction is $\sim 53\%$. This is partly because on average the stars have lower masses, and partly because there are fewer of them.}
\end{itemize}

%%%%%
\section*{acknowledgements}
The authors like to thank the anonymous referee for the comments that helped to significantly improve the paper. 
PFR, SW, SDC and AK acknowledge support via the European Research Council (ERC) starting grant No.
679852 `RADFEEDBACK'. 
DS and SW thank the Deutsche Forschungsgemeinschaft (DFG) for funding
via the SFB 956 `Conditions \& impact of star formation', via the sub-projects C5 and C6.
APW gratefully acknowledges the support of a consolidated
grant (ST/K00926/1) from the UK Science and Technology Facilities Council. 
The authors gratefully acknowledge the Gauss Centre for Supercomputing e.V. 
(www.gauss-centre.eu) for funding this project (ID: pr47pi) by providing computing time on the GCS Supercomputer SuperMUC at Leibniz Supercomputing Centre (www.lrz.de).
PFR acknowledges D. Price for providing the visualisation tool SPLASH \citep{Price11}.
%%%%%

%%%%%
\section*{data availability}
The data underlying this article will be shared on reasonable request to the corresponding author.
%%%%%

%%%%%
\bibliographystyle{mnras}
\bibliography{sample}
%%%%%

\label{lastpage}
\end{document}